\documentclass[aps,pra,superscriptaddress]{revtex4}

\usepackage{graphicx}
\usepackage{amsmath}
\usepackage{amssymb}
\usepackage[utf8]{inputenc}
\usepackage[english]{babel}

\usepackage{natbib}
\usepackage{color}
\usepackage{relsize}

\usepackage{bbm} 

\newcommand{\bra}[1]{\langle\,{#1}\, |}
\newcommand{\ket}[1]{|\,{#1}\,\rangle}

\newcommand{\oop}[1]{\bar{O}^{#1}_0}

\newlength{\mylenunit}
\usepackage{afterpage}

\begin{document}

\title{Long-range coherent energy transport in Photosystem II}

\author{Jan J.\ J.\ Roden}
\email{jan.roden@gmx.com}
\affiliation{Department of Chemistry, University of California, Berkeley, CA 94720, USA}

\author{Doran I.\ G.\ Bennett}
\affiliation{Department of Chemistry, University of California, Berkeley, CA 94720, USA}

\author{K.\ Birgitta Whaley}
\affiliation{Department of Chemistry, University of California, Berkeley, CA 94720, USA}

\date{\today}

\begin{abstract}
We simulate the long-range inter-complex electronic energy transfer in Photosystem II -- from the antenna complex, via a core complex, to the reaction center -- using a non-Markovian (ZOFE) quantum master equation description that allows us to quantify the electronic coherence involved in the energy transfer. 
We identify the pathways of the energy transfer in the network of coupled chromophores, using a description based on excitation probability currents.
We investigate how the energy transfer depends on the initial excitation -- localized, coherent initial excitation versus delocalized, incoherent initial excitation -- and find that the energy transfer is remarkably robust with respect to such strong variations of the initial condition.
To explore the importance of vibrationally enhanced transfer and to address the question of optimization in the system parameters, we vary the strength of the coupling between the electronic and the vibrational degrees of freedom. 
We find that the original parameters lie in a (broad) region that enables optimal transfer efficiency, and that the energy transfer appears to be very robust with respect to variations in the vibronic coupling.
Nevertheless, vibrationally enhanced transfer appears to be crucial to obtain a high transfer efficiency. 
We compare our quantum simulation to a ``classical'' rate equation based on a modified-Redfield/generalized-F\"{o}rster description that was previously used to simulate energy transfer dynamics in the entire Photosystem II complex, and find very good agreement between quantum and rate-equation simulation of the overall energy transfer dynamics.
\end{abstract}


\maketitle


\section{Introduction}

\begin{figure}
\centering
\includegraphics[width=0.8\mylenunit]{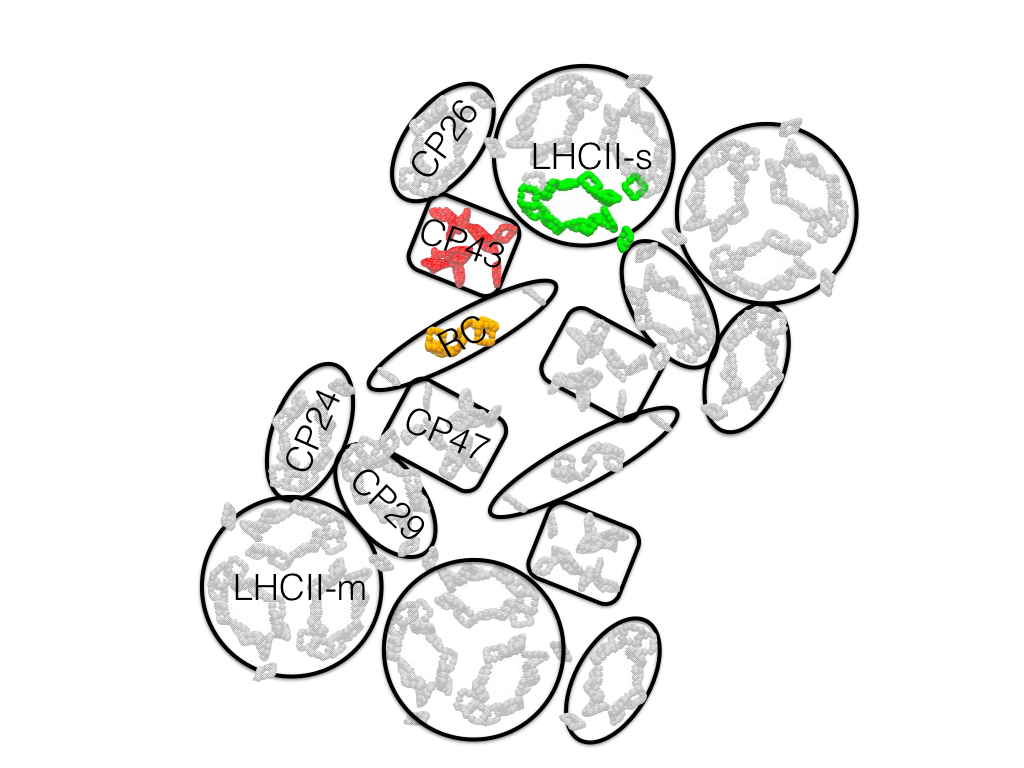}
\caption{Structural arrangement of the pigments within the PSII supercomplex, based on the structure determined by Caffarri and co-workers~\cite{caffarri2009_3052}. 
The protein scaffold is removed to show the pigments bound by each protein unit.
The pigments that belong to the LHCII-CP43-RC subsystem considered in the energy-transport simulations of the present paper are colored:
green in the LHCIIC monomer, red in CP43, and orange in the reaction center.
The other pigments not included in our truncated model are gray.  
}
\label{fig_structure_PSII}
\end{figure}

The initial steps in photosynthesis of green plants occur when Photosystem II (PSII) absorbs light and the nascent excitation is used to split water molecules into electrons, hydrogen ions, and molecular oxygen. 
The ensuing electrons and protons drive the subsequent photosynthetic reactions necessary for the production of adenosine tri-phosphate (ATP), the “chemical currency” of biological reactions~\cite{knowles1980_877}.
Greater than 80\% of all excitations initiated in PSII will result in productive photochemistry, causing the creation of an irreversibly separated electron-hole pair contributing to water splitting~\cite{caffarri2011_2094}. 
The mechanisms of efficient energy transport within PSII have been studied using spectroscopy~\cite{ruban2012_977, belgio2012_2761, marin2012_91, kruger2012_2669, kruger2010_3093, schlau2014single, novoderezhkin2011_681, novoderezhkin2011_17093, fuller2014_706, wells2014_11640, schlau2012_389, broess2006_3776} and numerical simulations~\cite{raszewski2008_4431, bennett2013_9164, kreisbeck2014_4045, shibata2013_6903}, with models based on spectroscopic data and structural information from X-ray crystal structures of the complexes~\cite{umena2011_55, liu2004_287, pan2011_309, muh2014_11848}. 
Recent work has focused on energy transfer within PSII supercomplexes, which represent the smallest photosynthetic unit that contains all of the relevant proteins for these first steps in photosynthesis~\cite{caffarri2011_2094, vanOort2010_922, caffarri2009_3052}. 
The largest PSII supercomplex isolated by Caffarri and co-workers~\cite{caffarri2009_3052} (called C$_{\rm 2}$S$_{\rm 2}$M$_{\rm 2}$) contains 326 pigments bound into an assembly of light harvesting proteins surrounding a PSII reaction center.
The C$_{\rm 2}$S$_{\rm 2}$M$_{\rm 2}$ supercomplex is a dimer, where each of the two monomers contains two antenna complexes (LHCII), three minor light harvesting complexes (CP24, CP26, and CP29), two core complexes (CP43, CP47), and one reaction center (RC). 
The arrangement of the complexes is shown in Figure~\ref{fig_structure_PSII}. 

Absorption of sunlight by chlorophyll pigments in the antenna complexes of PSII creates electronic excitations that are transferred to other pigments located in the same protein complex and from there to pigments in neighboring complexes.
In this way, the energy is transferred from the antenna complexes to the core complexes and from there to the reaction center. 
Because of these complicated dynamics, involving many degrees of freedom and a relatively large number of pigments, numerical simulations for PSII are very challenging.  
Consequently, simulations of energy transfer in PSII have been limited to smaller sub-complexes of PSII or to simpler rate-equation descriptions~\cite{raszewski2008_4431, bennett2013_9164, shibata2013_6903} rather than the full quantum dynamics simulations that have been performed for smaller light-harvesting systems such as the Fenna-Matthews-Olson complex~\cite{rebentrost2009_184102, ishizaki2009_17255, chin2010_065002, kreisbeck2011_2166, ritschel2011efficient, ritschel2011absence}.
Recently, in Ref.~\cite{bennett2013_9164}, simulations based on a combined modified-Redfield-generalized-F\"{o}rster (MRGF) rate equation approach were carried out for the entire PSII supercomplex, providing new insight into the interplay of short-range transfer dynamics inside the individual subcomplexes (proteins) and long-range inter-complex dynamics within a rate-equation kinetic analysis.

In recent years, spectroscopy experiments have provided evidence that there is coherence involved in excitonic energy transfer in light-harvesting complexes~\cite{engel2007evidence, romero2014_676}.
This observation raises questions about the role of such coherence in the energy transfer.
In particular, is coherence important for the efficiency of the transfer?
Is it relevant for design principles that aim to maximize this efficiency?
These questions have been addressed in many recent papers~\cite{olaya2008efficiency, scholes2011lessons, hoyer2012spatial, jesenko2013_174103, chenu2014dynamic, scholes2012_9374, kassal2013_362, rebentrost2009_9942, hoyer2010limits, shim2012atomistic, abramavicius2010quantum, sarovar2013design, chenu2015coherence}.
In a companion paper~\cite{roden2015prob_current} of the present paper, the role of electronic coherence in electronic excitation energy transfer is analyzed in the framework of excitation probability currents.

In the present paper, we investigate the energy transfer dynamics in PSII by means of a full quantum simulation, using a non-Markovian quantum master equation description.
This not only allows us to quantify the electronic coherence involved in the energy transfer over length scales including several subcomplexes, but also to calculate its contribution to the energy transfer in terms of excitation probability currents, using the framework of Ref.~\cite{roden2015prob_current}.
The analysis of the probability currents also provides insight into the pathways of the energy transfer in the large network of coupled pigments. 
Using a non-Markovian quantum master equation allows us to account for the non-negligible memory times, i.e., the non-Markovianity, associated with both the intra-molecular vibrations of the pigments and the vibrations of the protein environment of the pigments to which the electronic degrees of freedom couple.
Specifically, we investigate the long-range inter-complex energy transfer from the antenna complex via a core complex to the reaction center.
To this end, we simulate the energy transfer in a subsystem containing a LHCII monomer, the core complex CP43, and the reaction center (see Figure~\ref{fig_structure_PSII}).
This requires incorporation of $\sim 30$ pigments in the simulation. 
To be able to efficiently treat non-Markovian dynamics for this number of pigments, we use the non-Markovian ZOFE quantum master equation that has been developed in recent years~\cite{strunz2004convolutionless}, successfully applied to biological light-harvesting complexes, and also tested against exact results~\cite{ritschel2011efficient, ritschel2011absence}.
This approach allows calculations for wide parameter ranges of the model. 
To compare the simulation results of the ZOFE quantum master equation and the MRGF rate equation of Ref.~\cite{bennett2013_9164}, we analyze here the results of both methods for excitonic energy transfer in the subsystem of PSII described above.

There is an ongoing discussion about whether and how energy transfer in light-harvesting systems depends on the initial conditions -- in particular, initial excitation in laser spectroscopy experiments can be coherent and localized, whereas in natural initial conditions through absorption of sunlight, delocalized, incoherent states are often assumed~\cite{kassal2013_362, jesenko2013_174103, han2013_8199, dorfman2013_2746}.
Motivated by these discussions, in the present paper we simulate the energy transfer dynamics for very different initial states to ascertain how long-range energy transfer in PSII depends on the initial condition.
We consider first initial excitation that is localized in the antenna complex, then look at how the energy transfer changes when all pigments in the antenna complex, core complex, and the reaction center carry the initial electronic excitation, i.e., when the initial state is completely delocalized over all three complexes.
We also analyze the role of vibrations -- in particular, intra-molecular vibrations of the pigments and vibrations of the protein environment -- that couple to the electronic degrees of freedom and that can enhance the electronic energy transfer through vibrationally enhanced transport, i.e., by creating resonances between pigments through vibronic energy levels of the pigments.
In the present work, we vary the coupling between the electronic and the vibrational degrees of freedom to investigate how the energy transfer depends on this coupling to the vibrations. 

The remainder of the paper is structured as follows.
In Section~\ref{sec_model}, we describe the model that we use for the simulation of the energy transfer in PSII.
The simulated energy transfer dynamics are then shown in Section~\ref{sec_simul_results}, for an initial state in which the initial excitation is localized in the antenna complex.
In Section~\ref{sec_robustness} we compare to a simulation where the initial excitation is delocalized over all three complexes considered here.
In Section~\ref{sec_robustness}, the dependency of the energy transfer on the coupling between the electronic and vibrational degrees of freedom is investigated. 
A summary of the results and concluding remarks are given in Section~\ref{sec_conclusion}.

\section{Model}
\label{sec_model}

We aim to investigate the long-range transport of excitation energy in the Photosystem II supercomplex, from the surrounding antenna complexes, through the outer components of the supercomplex, to the reaction center -- by means of a full quantum simulation. 
To this end, we model a subsection of the Photosystem II supercomplex originally isolated by Croce, et al.~\cite{caffarri2009_3052} and subsequently used as a foundation for a structure based excitation energy transport model to explain the measured fluorescence lifetimes~\cite{caffarri2011_2094}. 
Figure~\ref{fig_structure_PSII} shows the largest PSII supercomplex previously modeled~\cite{caffarri2011_2094, bennett2013_9164}. 
The colored pigments in LHCII, CP43, and the reaction center shown in Figure~\ref{fig_structure_PSII} represent the subsystem that we model in this work:
it contains 33 chromophores distributed between a LHCII monomer (14 pigments), CP43 (13 pigments), and a truncated reation center (6 pigments).

\subsection{System parameters} 
\label{sec_sys_params}

\subsubsection{Electronic degrees of freedom}

\begin{figure}
\centering
\includegraphics[width=0.5\mylenunit]{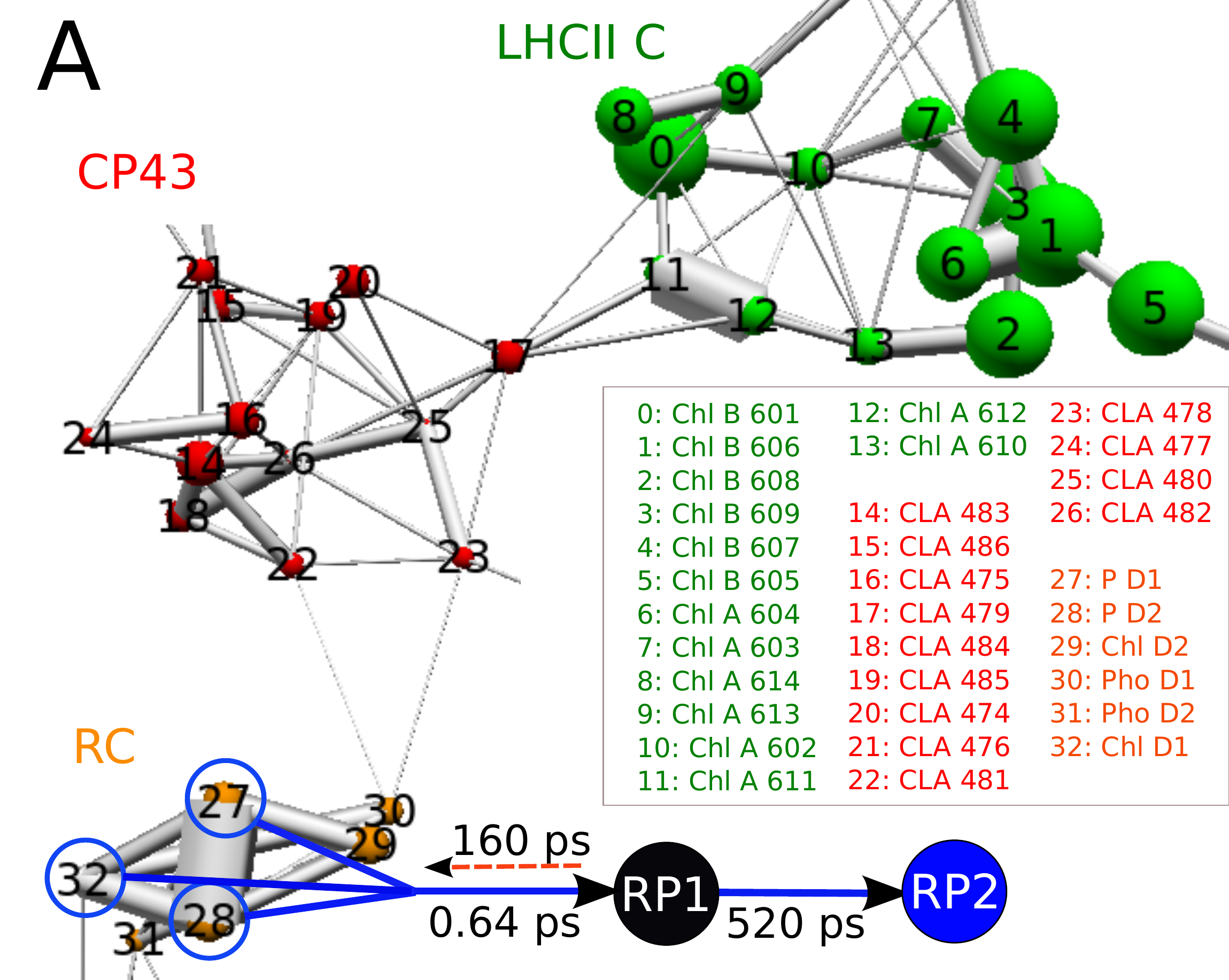}
\includegraphics[width=0.45\mylenunit]{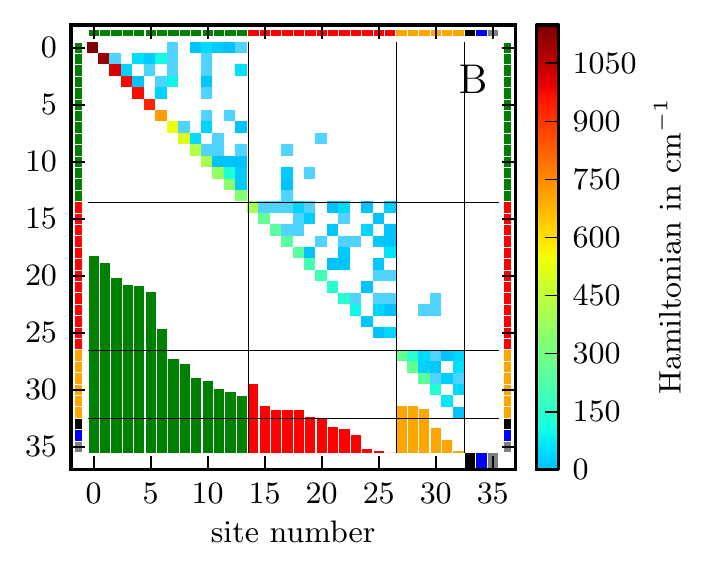}
\caption{Electronic part of the model used for the simulation of energy transport from the LHCII antenna complex via the CP43 core complex to the reaction center (only monomer C of LHCII is included). 
This subsystem of PSII contains in total 33 pigments.
Also included are the two radical pair states RP1 and RP2 of Ref.~\cite{bennett2013_9164} with the characteristic transfer times for the description of the charge transfer.
{\bf Figure~A:} spatial and energetic structure. 
The spheres are located at the positions of the pigments and their sizes show the relative electronic transition energies of the pigments.
The numbering of the pigments starts with pigment~0 in LHCII that has the highest energy (15,890~cm$^{-1}$), then the number increases with decreasing energy within each protein;
pigment~32 has the lowest energy (14,740~cm$^{-1}$). 
The connection lines between the spheres represent the electronic interactions between the pigments; their thickness is proportional to the strength of the respective interaction (absolute values);
the maximum interaction strength is 150 cm$^{-1}$.
Transfer of population between pigments 27, 28, and 32 and RP1, and from RP1 to RP2 is depicted by arrows.
In addition to the numbering of the pigments used in the present work, the common names of the pigments used in Ref.~\cite{bennett2013_9164} and references therein are given in the listing.
{\bf Figure~B:} corresponding Hamiltonian matrix in the site (pigment) basis, where the relative transition energies of Figure~A are on the diagonal and the absolute values of the interactions are the off-diagonal elements.
For better visibility, the {\bf interactions are increased by a factor of ten with respect to their true values}.
The coloring at the edges indicates to which protein the sites belong:
Green: LHCII C.
Red: CP43.
Orange: RC.
Black: RP1 state.
Blue: RP2 state.
Gray: ground state.
All parameters for this model are taken from Ref.~\cite{bennett2013_9164}.
}
\label{fig_structure_and_ham}
\end{figure}

Our model contains 33 chromophores in total, each of which we model as having two electronic states separated by the local excitation energy, referred to here as the 'site energy', which varies across different pigments depending upon their interactions with the local protein environment. 
We further describe the interaction between the chromophores by electronic matrix elements that couple electronic excitation of a chromophore to electronic excitation of other chromophores. 
In the subspace of single excitations that is relevant for energy transport, this is described by the electronic Hamiltonian, 
\begin{equation}
  H_{\rm elec} = \sum_n \varepsilon_n\ket{n}\bra{n} + \sum_{nm}V_{n,m}\ket{n}\bra{m}
\end{equation}
containing $N$ site energies ($\varepsilon_n$) and the $N\times N$ coupling matrix $V$. 
Here, $\ket{n}$ is the state in which only pigment $n$ is excited and all others are in the ground state.
The construction of an electronic Hamiltonian for the PSII supercomplex has been described previously~\cite{bennett2013_9164}. 
The couplings between pigments contained within the same protein have been extracted from the literature for each complex~\cite{bennett2013_9164}, where they were constructed to reproduce the available linear and non-linear spectroscopy data for ensembles of the isolated pigment-protein complex. 
The coupling between pigments in different proteins has been calculated assuming a dipole-dipole coupling~\cite{bennett2013_9164}. 
All energies and coupling parameters are taken from Ref.~\cite{bennett2013_9164}, and are shown in Figure~\ref{fig_structure_and_ham}, together with the structure of the LHCII-CP43-RC truncated supercomplex.

\subsubsection{Coupling to vibrations} 

The electronic states of each chromophore are coupled to a large collection of vibrational modes that represent intra-molecular vibrational modes as well as vibrational modes of the surrounding protein scaffold. 
In the extreme limit of low-frequency vibrations, the long-time-scale protein conformation dynamics give rise to an inhomogeneous distribution of energies for each chromophore within any ensemble measurement. 
In contrast to Ref.~\cite{bennett2013_9164}, however, where ensemble averaging over static disorder in the site energies is performed, in the present work we do not describe ensemble averaging and instead use the electronic Hamiltonian with average site energies for our simulations. 
While ensemble averaging can be critical for comparing to experimental measurements, in this paper we investigate quantities that are primarily focused on the fundamental process of excitation energy transport within an individual PSII supercomplex. 
Here we may take the average energies instead of random energies of a single realization to create a ``best-case scenario'' for transport efficiency and emergence of coherence.

We describe here the intra-molecular vibrational modes of the pigments together with the vibrational modes of the protein environment by the single vibrational Hamiltonian
\begin{equation}
H_{\rm vib} = \sum_n \sum_{\lambda} \omega_{n\lambda}a^{\dagger}_{n\lambda} a_{n\lambda},
\end{equation}
where we approximate the vibrational modes to be harmonic.
Here, $a_{n\lambda}$ is the annihilation operator of mode $\lambda$ with frequency $\omega_{n\lambda}$ that belongs to a pigment $n$. 
(Here and throughout the paper we set $\hbar=1$). 
In this description, each pigment has its own separate bath of vibrational modes that does not directly couple to the modes of another pigment. 
Indirectly, however, such a coupling can come about through the electronic dipole-dipole interaction between two pigments. 

We assume that electronic excitation of a pigment couples linearly to its own vibrational modes, such that the overall coupling between the electronic degrees of freedom and the vibrations is given by
\begin{equation}
  H_{\rm elec-vib} = \sum_n L_n \sum_{\lambda}\kappa_{n\lambda}(a^{\dagger}_{n\lambda} + a_{n\lambda}),
\end{equation}
where the system operators $L_n$ are the projectors $L_n=-\ket{n}\bra{n}$. 
The coupling constants $\kappa_{n\lambda}$ that describe the strength of the coupling of electronic excitation of pigment $n$ to vibrational mode $\lambda$ of this pigment, can be expressed through the vibrational spectral density of pigment $n$
\begin{equation}
\label{spec_dens}
  J_n(\omega) = \sum_{\lambda}|\kappa_{n\lambda}|^2 \ \delta(\omega - \omega_{n\lambda}).
\end{equation}
Assuming that the spectral density is a continuous function, this leads to the vibrational correlation function
\begin{equation}
  \alpha_n(\tau) = \int_0^{\infty} d\omega\ J_n(\omega)\left[{\rm coth}\left(\frac{\omega}{2kT}\right)\cos(\omega\tau) - i\sin(\omega\tau)\right],
\end{equation}
which depends on the temperature $T$. 
The Fourier transform
\begin{equation}
\label{c_of_omega}
  C_n(\omega) = \int_{-\infty}^{\infty} d\tau \ e^{i\omega \tau} \alpha_n(\tau),
\end{equation}
which we shall refer to as the 'vibrational correlation spectrum', will be useful later on for application of the ZOFE master equation.
It can also be directly calculated from the spectral density $J_n(\omega)$ (or vice versa) via the relation~\cite{mukamel1982principles}
\begin{equation}
\label{c_of_omega_from_j_of_omega}
  C_n(\omega) = 2\left[1 + \left(e^{\omega / (kT)} - 1\right)^{-1}\right]J_n^A(\omega)
\end{equation}
with the anti-symmetrized spectral density
\begin{equation}
J_n^A(\omega) = \left\{
\begin{array}{lcl}
J_n(\omega) &; & \omega \ge 0
\\
-J_n(-\omega) &;& \omega <0
\end{array}
\right.
\end{equation}
Peaks in $C_n(\omega)$ that are narrow compared to the relevant energies of the electronic degrees of freedom will lead to non-Markovian dynamics. 
When we apply the ZOFE master equation below, we will fit $C_n(\omega)$ using a sum of Lorentzians.

\subsubsection{Charge transfer in the reaction center} 

Up to this point, we have described a system where electronic excitations of individual pigments are coupled both to other pigments and to vibrational modes. 
Two other key features need to be incorporated into our model to make it a reasonable model of the initial stages of photosynthesis in PSII.
These are, i) energy capture via charge separation at the reaction center, and ii) loss of excitation via non-radiative or fluorescence processes. 

We include these dissipative processes by extending our description of the electronic degrees of freedom to include the phenomenological radical pair states ($\ket{\rm RP1}$ and $\ket{\rm RP2}$), as described in Ref.~\cite{bennett2013_9164}, and the ground state $\ket{0}$, where all pigments are in their ground electronic states. 

In this truncated supercomplex model, energy is transferred from the excited states of three of the pigments in the reaction center to the first radical pair state RP1. 
On a slower time scale, some of this excitation is transported back to the excited states of the three pigments in the reaction center. 
From RP1, the energy is irreversibly and more slowly transferred to the second radical pair state RP2.
The latter is the last step in our model and describes the irreversible trapping. 
The characteristic transfer times that are found in Ref.~\cite{bennett2013_9164} are 0.64~ps from the excited states of the RC pigments to RP1, 160 ps from RP1 back to the excited states of the RC pigments, and 520 ps for the irreversible transfer from RP1 to RP2. 
Electronic excitations present on any pigment are assumed to have equal probability of being lost through either fluorescent or non-radiative decay mechanisms, with a characteristic decay time of 1.78~ns (combining the decay times of 2~ns for non-radiative decay and 16~ns for fluorescence used in Ref.~\cite{bennett2013_9164}). 

The Hamiltonian term describing the radical pair states and the ground state is given by 
\begin{equation}
  H_{\rm RP+GS} = \varepsilon_{\rm GS}\ket{\rm 0}\bra{\rm 0} + \varepsilon_{\rm RP1}\ket{\rm RP1}\bra{\rm RP1} + \varepsilon_{\rm RP2}\ket{\rm RP2}\bra{\rm RP2},
\end{equation}
where $\varepsilon_{\rm GS}$ is the ground state energy, and $\varepsilon_{\rm RP1}$ and $\varepsilon_{\rm RP2}$ are the energies of the RP1 and RP2 state.

We shall describe in detail the dynamical model employed to treat excitation transfer into either the radical pair or ground state of the pigment in Section~\ref{sec_zofe_lindblad_master}.

\subsection{Excitation energy transport}

Based on the previous section (Section~\ref{sec_sys_params}), the total Hamiltonian of our model is then
\begin{equation}
 H_{\rm tot} = H_{\rm elec} + H_{\rm RP+GS} + H_{\rm elec-vib} + H_{\rm vib}.
\end{equation}
Since this Hamiltonian comprises a large number of degrees of freedom, direct calculation of the excitation energy transport by explicitly solving the full quantum dynamics is not realistic.
Therefore, we use instead the framework of the description of open quantum systems, and divide the total Hamiltonian into three components:
a system ($H_{\rm sys}$), an environment ($H_{\rm env}$), and the interaction between system and environment ($H_{\rm sys-env}$).
The system part $H_{\rm sys}$ should contain the quantities relevant to the energy transport, that is, most importantly the probabilities of electronic excitation of the different pigments and the excitation of the radical pair states in the reaction center.
On the other hand, we want to keep $H_{\rm sys}$ as small as possible to keep the calculation numerically manageable.
Therefore, we choose this to consist only of the electronic degrees of freedom of the pigments and the radical pair states.
The vibrations are then treated as the environment.
That is, we have
\begin{equation}
  H_{\rm sys} \equiv H_{\rm elec} + H_{\rm RP+GS}, \ \ \ \ \ H_{\rm env} \equiv H_{\rm vib}, \ \ \ \ \ \mbox{and} \ \ H_{\rm sys-env} \equiv H_{\rm elec-vib}.
\end{equation}

The excitation energy transport and the electronic coherence can be extracted from the reduced density matrix of the system part, i.e.,
\begin{equation}
  \rho(t) = {\rm Tr}_{\rm env} \ \rho_{\rm tot}(t),
\end{equation}
which is obtained from the total density matrix by tracing out the degrees of freedom associated with the environment, i.e., the vibrations. 
In the basis of the states $\ket{n}$ of the local excitations of the pigments, the diagonal of the reduced density matrix gives the populations of the excited electronic states of the pigments, the radical pair states, and the ground state. 
The off-diagonal elements are the electronic coherences between the pigments, which constitute a major focus of the present work.

\subsubsection{ZOFE master equation}

For the simulations of the dynamics of energy transport and coherence we use the ZOFE master equation, which allows us to take into account the non-Markovian effects in the coupling between the electronic degrees of freedom and the vibrational environment.
The ZOFE master equation is given by~\cite{strunz2004convolutionless, ritschel2011efficient}
\begin{equation}
  \label{zofe_master}
  \partial_t \rho(t) = -i\left[H_{\rm sys}, \rho\right] + \sum_n\left(L_n\rho\oop{(n)\dagger}  + \oop{(n)}\rho L_n^{\dagger} - L_n^{\dagger}\oop{(n)}\rho - \rho\oop{(n)\dagger}L_n \right)
\end{equation}
where the auxiliary operator
\begin{equation}
  \label{o_bar}
  \oop{(n)}(t)=\int_0^t ds \, \alpha_n(t-s) O^{(n)}_0(t,s),
\end{equation}
captures the non-Markovian effects, and with initial condition 
\begin{equation}
  \label{o_init}
  O^{(n)}_0(t,t)=L_n.
\end{equation}
Depending on the form of the environment correlation function $\alpha_n(t-s)$, a closed evolution equation for the auxiliary operator $\oop{(n)}(t)$ can be obtained~\cite{yu1999_91, strunz2004convolutionless, ritschel2011efficient}.
For an environment correlation function that is a sum of damped oscillating terms,
\begin{equation}
\label{alpha_sum_exps}
  \alpha_n(t-s) = \sum_j\Gamma_{nj}e^{-i\Omega_{nj}(t-s)-\gamma_{nj}|t-s|},
\end{equation} 
such a closed evolution equation can be found~\cite{ritschel2011efficient}.
Writing the operator $\oop{(n)}(t)$ as the sum $\oop{(n)}=\sum_j \oop{(nj)}$, the closed evolution equation to obtain the operators $\oop{(nj)}(t)$ is given by~\cite{ritschel2011efficient}
\begin{equation}
  \label{oop_evol}
  \partial_t\oop{(nj)}(t) = \Gamma_{nj}L_n - (i\Omega_{nj}+\gamma_{nj})\oop{(nj)} - \left[iH_{\rm sys} + \sum_m L_m^{\dagger}\oop{(m)}\ ,\ \oop{(nj)}\right],
\end{equation}
with initial condition $\oop{(nj)}(t=0)=0$ and $O^{(n)}_0(t,t)=L_n$.
In the present work, we use these evolution equations with coupling operators $L_n=-\ket{n}\bra{n}$ for each pigment $n$.

\subsubsection{Approximate environment spectral density for ZOFE master equation} 
\label{sec_approx_spec_dens_for_zofe}

\begin{figure}
\centering
\includegraphics[width=0.49\mylenunit]{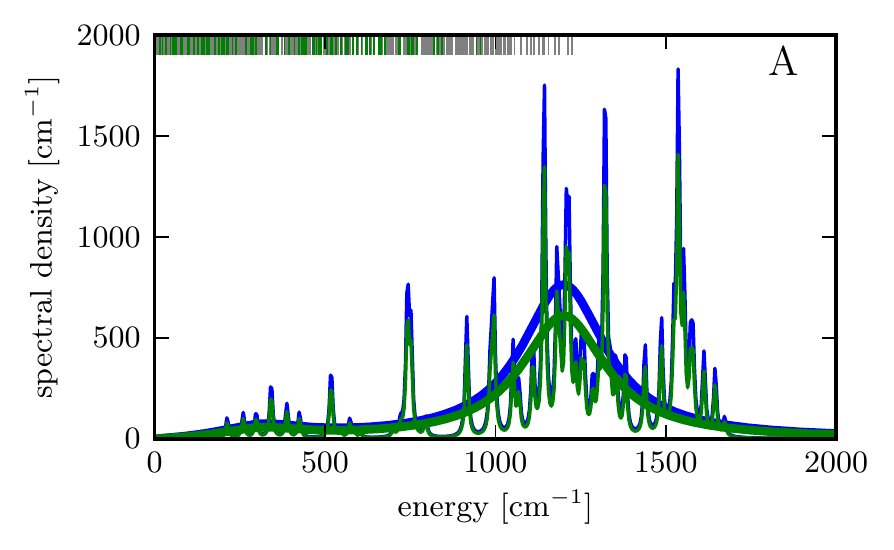}
\includegraphics[width=0.49\mylenunit]{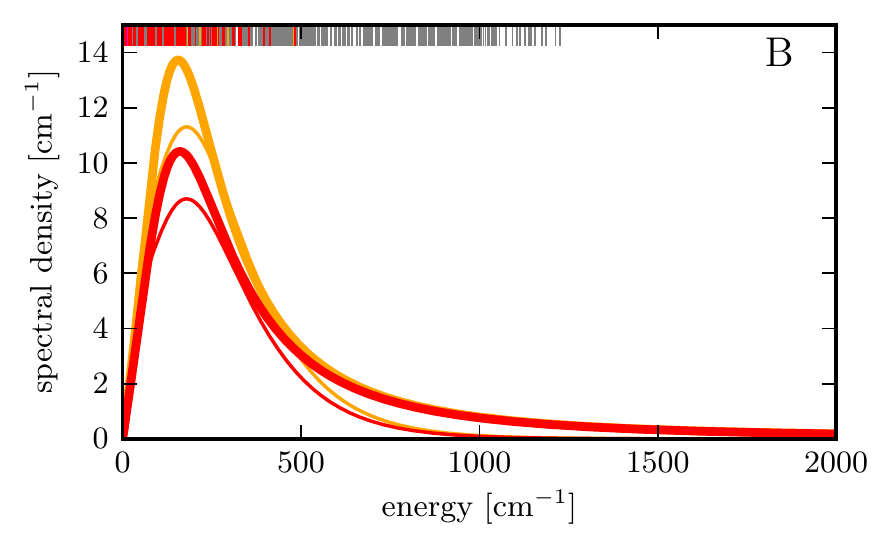}
\caption{Spectral densities of intramolecular and protein vibrations that the different pigments in the different proteins couple to.
The thin gray bars at the upper edge show where all transition energies of the electronic system Hamiltonian are, that is, all energy differences between the eigenenergies of the Hamiltonian.
{\bf Figure~A:} The blue and green narrow peaks are the two different spectral densities of the Chl B (blue) and Chl A (green) pigments in the LHCII antenna, respectively, from Ref.~\cite{bennett2013_9164} and references therein.
The broad blue and green peaks are the corresponding fitted spectral densities that are used in the ZOFE simulation.
The green bars at the upper edge show the transition energies between the exciton states localized in the LHCII antenna.
{\bf Figure~B:} The thin red and orange curves show the spectral densities of the pigments in CP43 (red) and RC (orange), respectively, from Ref.~\cite{bennett2013_9164} and references therein.
The thick curves are the corresponding fitted spectral densities for the ZOFE simulation.
Red, magenta, and orange bars at the upper edge indicate the transition energies between exciton states localized in CP43, localized in both, CP43 and RC, and localized in RC, respectively. 
We note that the spectral densities in Figure~A are much larger than the ones in Figure~B (note range of y-axis).  
}
\label{fig_spectral_density}
\end{figure}

The form of $\alpha_n(t-s)$ in Equation~(\ref{alpha_sum_exps}) corresponds to a sum of Lorentzians centered at frequencies $\Omega_{nj}$ and with weights $\Gamma_{nj}$ and widths $\gamma_{nj}$ in the environment correlation spectrum $C_n(\omega)$ (the Fourier transform of $\alpha_n(t-s)$):
\begin{equation}
\label{c_of_omega_lorentzians}
  C_n(\omega) = \frac{2}{\pi}\sum_j \Gamma_{nj} \frac{\gamma_{nj}}{(\omega - \Omega_{nj})^2 + \gamma_{nj}^2}.
\end{equation} 
We use this expression to fit the environment correlation spectra $C_n(\omega)$ of the pigments to experiment-based spectra, in order to obtain the parameters $\Omega_{nj}$, $\Gamma_{nj}$, and $\gamma_{nj}$ that we need when we solve the evolution equation for the auxiliary operator Eq.~(\ref{oop_evol}).
The resulting spectral densities $J_n(\omega)$ (Eq.~(\ref{spec_dens})) that we use in the ZOFE simulation are shown in Figure~\ref{fig_spectral_density}, together with the original, measurement-based spectral densities of Ref.~\cite{bennett2013_9164}.
As one can see in Figure~\ref{fig_spectral_density}A, the fitted spectral density that we use for the LHCII pigments in the ZOFE simulation is very broad compared to the narrow peaks of the original spectral density. 
We approximate the original spectral density with these broad peaks, because the ZOFE approximation that we use in the simulation is quite limited in the degree of non-Markovianity that it can handle.
Thus, the memory time of the vibrations can not be too long compared to the time scale of the electronic dynamics. 
Narrow peaks would result in a long memory time, so we are limited to broad peaks that capture only the general features of the coupling to the vibrations. 

The approximate spectral densites in Figure~\ref{fig_spectral_density}A also neglect the high-energy contributions of the original spectral density around and above 1,500~cm$^{-1}$.
Since these peaks are off-resonant with the system energies, which are marked by the bars at the upper edge, we assume that they do not have a large influence on the system dynamics and it is reasonable to neglect them.
In Ref.~\cite{ritschel2011efficient}, it was shown that the influence of off-resonant parts of the spectral density on excitation energy transport is indeed negligible.  

In Figure~\ref{fig_spectral_density}B, the spectral densities for CP43 and the RC that we use for the ZOFE simulation are shown.
They are only in rough agreement with the original, measurement-based spectral densities from Ref.~\cite{bennett2013_9164}.
This difference arises because these spectral densities are obtained not by directly fitting the spectral densities $J_n(\omega)$, but by fitting the environment correlation spectra $C_n(\omega)$ with Lorentzians, according to Equation~(\ref{c_of_omega_lorentzians}).
To summarize, we apply the following procedure:
\begin{enumerate}

\item Calculate the environment correlation spectra $C_n(\omega)$ from the experimentally determined spectral densities $J_n(\omega)$ from Ref.~\cite{bennett2013_9164}, using Equation~(\ref{c_of_omega_from_j_of_omega}).
For this step, we need to assume a temperature that enters in $C_n(\omega)$.
Since we want to simulate energy transport at room temperature, we choose $T=300$~K.

\item Fit these $C_n(\omega)$ with Lorentzians according to Equation~(\ref{c_of_omega_lorentzians}), to determine the parameters for the evolution equations~(\ref{oop_evol}) for the ZOFE auxiliary operators. 
In this work, we use only two Lorentzians for the LHCII pigments and one Lorentzian for the CP43 and RC pigments, to keep the corresponding number of evolution equatons small (see index $j$ in Eq.~(\ref{oop_evol})).

\end{enumerate}
As a consistency check, we convert these fitted sums of Lorentzians $C_n(\omega)$ back to spectral densities $J_n(\omega)$ to compare with the original spectral densities, as is done in Figure~\ref{fig_spectral_density}.

\subsubsection{ZOFE master equation: excitation loss and radical pair states}
\label{sec_zofe_lindblad_master}

To simulate the transfer of energy to the radical pair states in the reaction center, and to take radiative and non-radiative decay of the excited electronic states of the pigments into account, we add Lindblad-master-equation terms to the ZOFE master equation. 
An analogous treatment of radiative decay and trapping in the reaction center is employed in Ref.~\cite{kreisbeck2011_2166}, where a non-Markovian master equation is also extended by corresponding Lindblad terms, for the simulation of energy transport in the Fenna-Matthews-Olson complex.
(See also Refs.~\cite{caruso2009_105106, jesenko2013_174103} for related Lindblad models for trapping and decay.)

This Markovian Lindblad description of the trapping in the reaction center is a rough approximation of the full charge separation dynamics, which we use here in order to create a minimal model to capture the primary dynamics. 
We assume that the characteristic transfer times to the radical pair states that are found in Ref.~\cite{bennett2013_9164} give a reasonable approximation and therefore describe the trapping part of our simulation with Lindblad terms based on these characteristic transfer times.

In the Markov limit, where the environment correlation function $\alpha(\tau)$ decays fast enough compared to all relevant time scales of the dynamics, the ZOFE master equation itself becomes a Lindblad master equation~\cite{strunz2004convolutionless}.
It is therefore consistent to add Lindblad terms to the ZOFE master equation, since they are equivalent to additional ZOFE master equation terms with fast decaying correlation functions.   
In the Markov limit $\alpha_i(\tau) = \widetilde{\gamma}_i\, \delta(\tau)$, the auxiliary operator of the ZOFE master equation becomes the constant operator $\oop{(i)}(t)=\frac{1}{2}\widetilde{\gamma}_i \widetilde{L}_i$ (see Eqs.~(\ref{o_bar}) and (\ref{o_init}))~\cite{strunz2004convolutionless}.
(Here, we have replaced the index $n$, which is used above to refer to the pigments, by a more general index $i$, to refer to the respective process described by a system operator $\widetilde{L}_i$ and a corresponding coupling constant $\widetilde{\gamma}_i$; 
e.g.\ relaxation from a state $\ket{f}$ to a state $\ket{t}$ is described by $\widetilde{L}_{ft}=\ket{t}\bra{f}$.)  
Inserting this constant auxiliary operator into the ZOFE master equation~(\ref{zofe_master}) gives the well-known Lindblad master equation
\begin{equation}
  \label{lindblad_master}
  \partial_t \rho(t) = -i\left[H_{\rm sys}, \rho\right] 
  + \sum_i\widetilde{\gamma}_i\left(\widetilde{L}_i\rho \widetilde{L}_i^{\dagger} - \frac{1}{2}\widetilde{L}_i^{\dagger}\widetilde{L}_i\rho - \frac{1}{2}\rho \widetilde{L}_i^{\dagger}\widetilde{L}_i \right),
\end{equation}
with coupling (Lindblad) operators $\widetilde{L}_i$ and corresponding coupling constants $\widetilde{\gamma}_i$.
We use such a Lindblad description for the following processes:
\begin{enumerate}
  
\item Radiative and non-radiative decay of the electronic excitation of all pigments ($n = 0\dots 32$) is described through Lindblad operators $\widetilde{L}_n^{\rm decay}=\ket{\rm 0}\bra{n}$ and the corresponding coupling constant $\widetilde{\gamma}_n^{\rm decay}$ is given by the decay rate.
Here, $\ket{\rm 0}$ denotes the electronic ground state.
The decay rate is chosen such that it corresponds to a characteristic decay time of 1.78~ns (combining the two decay times of 2~ns for non-radiative decay and 16~ns for fluorescent decay assumed in Ref.~\cite{bennett2013_9164}).
  
\item Transfer of excitation from the reaction center pigments 27, 28, and 32 to radical pair state RP1 through $\widetilde{L}_n^{\rm RP1}=\ket{\rm RP1}\bra{n}$ with rates $\widetilde{\gamma}_n^{\rm RP1}$ (where $n=27,28,32$).
From all three pigments the transfer to RP1 is assumed to occur with the same rate, which is set at the transfer time of 0.64 ps found in Ref.~\cite{bennett2013_9164}.
  
\item Transfer from RP1 back to the reaction center pigments 27, 28, 32 through $\widetilde{L}_n^{\rm RC}=\ket{n}\bra{\rm RP1}$ with rates $\widetilde{\gamma}_n^{\rm RC}$.
It is assumed that equal amounts of population are transferred back to the three pigments, so that each of the three rates is one third of the overall rate corresponding to the characteristic transfer time of 160 ps found in Ref.~\cite{bennett2013_9164}.
  \item Irreversible transfer from RP1 to radical pair state RP2 through $\widetilde{L}^{\rm RP2}=\ket{\rm RP2}\bra{\rm RP1}$ with a rate $\widetilde{\gamma}^{\rm RP2}$.
This rate is chosen according to the characteristic transfer time of 520 ps found in Ref.~\cite{bennett2013_9164}.
\end{enumerate}

For each of these processes, a corresponding Lindblad term is added to the ZOFE master equation, so that the complete master equation we solve in our simulations is
\begin{equation}
\label{zofe_lindblad_master}
\begin{split}
  \partial_t \rho(t) = -i\left[H_{\rm sys}, \rho\right] & + \sum_{n=0}^{32}\left(L_n\rho\oop{(n)\dagger}  + \oop{(n)}\rho L_n^{\dagger} - L_n^{\dagger}\oop{(n)}\rho - \rho\oop{(n)\dagger}L_n \right)\\
&  + \sum_{i=0}^{39}\widetilde{\gamma}_i\left(\widetilde{L}_i\rho \widetilde{L}_i^{\dagger} - \frac{1}{2}\widetilde{L}_i^{\dagger}\widetilde{L}_i\rho - \frac{1}{2}\rho \widetilde{L}_i^{\dagger}\widetilde{L}_i \right),
\end{split}
\end{equation}
where the Lindblad operators $\widetilde{L}_i$ and corresponding coupling constants (rates) $\widetilde{\gamma}_i$ in the Lindblad term in the second line of the equation belong to the respective processes that are listed above, and the index $i$ of the sum refers to the individual processes.
In the ZOFE term in the first line, the ZOFE coupling operators $L_n=-\ket{n}\bra{n}$ couple only electronic excitation of the pigments to the non-Markovian vibrational environment. 
The radical pair states $\ket{\rm RP1}$ and $\ket{\rm RP2}$ are not coupled to the non-Markovian vibrations. 

The system density matrix $\rho(t)$ now has components for the states $\ket{n}$, in which pigment $n$ is excited, as well as components for the radical pair states $\ket{\rm RP1}$ and $\ket{\rm RP2}$ and for the ground state $\ket{\rm 0}$.
$\rho(t)$ is obtained by solving this master equation together with the evolution equation for the ZOFE auxiliary operators.
For the truncated supercomplex considered here, $\rho(t)$ is then a 36 dimensional square matrix.
The system Hamiltonian does not contain coupling to or between RP1, RP2, and the ground state; 
all coupling to and between these states arises through the Lindblad terms.

\subsubsection{Population currents and the contribution of coherence}

The transfer of excitation energy between the pigments can be expressed in terms of currents of excitation probability between the pigments. 
As shown in a companion paper~\cite{roden2015prob_current}, this population current description straightforwardly identifies the contribution of coherence to the energy transfer, and this description also clearly separates the respective contributions of unitary dynamics, dephasing, and relaxation to the population currents between the pigments~\cite{roden2015prob_current}.
Another advantage of the population-current description is that it clearly shows the individual pathways that the energy transfer takes between the pigments. 
By integrating the currents over time, it is then possible to see the respective amounts of population that have been transported via each pathway.
This insight is useful to identify and analyze the functionality of the light-harvesting supercomplex and its constituent complexes and pigments.

Because the total electronic excitation probability is conserved across the pigments -- that is, the reduced electronic density matrix fulfills $\sum_n \rho_{nn}(t) = 1$ at all times -- a continuity equation~\cite{roden2015prob_current}
\begin{equation}
\label{conti_eq}
\partial_t\rho_{nn}(t) = \sum_{m\neq n}j_{mn}(t)
\end{equation}
holds, where $j_{mn}(t)$ is the (net) population current at time $t$ that transports population (i.e., excitation probability) from a pigment~$m$ to another pigment~$n$. 
When $j_{mn}(t)$ is positive, excitation is transported from pigment~$m$ to pigment~$n$;
when it is negative, the energy transfer goes in the other direction, i.e., from pigment~$n$ to pigment~$m$. 

Based on the continuity equation~(\ref{conti_eq}), for quantum master equations of the form used in the present work Eq.~(\ref{zofe_lindblad_master}), it is shown in Ref.~\cite{roden2015prob_current} that the individual population currents $j_{mn}(t)$ between the pigments can be calculated through
\begin{alignat}{3}
\label{total_current}
  j_{mn}(t) & =  \ j_{mn}^{\rm unitary}(t)            & {} + {} & \ \ j_{mn}^{\rm dephas}(t)\ & {} + {} & \ \ j_{mn}^{\rm relax}(t)\nonumber \\[5pt]
           & =  \ 2 V_{mn}{\rm Im}(\rho_{mn}(t))\  & {} + {} & \ \ 0                   & {} + {} & \ \ \left(\gamma_{mn}\rho_{mm}(t) - \gamma_{nm}\rho_{nn}(t)\right),
\end{alignat}
where $j_{mn}^{\rm unitary}$, $j_{mn}^{\rm dephas}$, and $j_{mn}^{\rm relax}$ are the respective contributions of unitary dynamics, dephasing, and electronic relaxation, which are specified in the second line of the equation.
Here, $\rho_{mn}(t)$ are the coherences between the sites, $\rho_{nn}(t)$ are the populations of the sites, and $V_{mn}$ are the inter-site couplings and $\gamma_{mn}$ is the rate of electronic relaxation from a site (or state) $m$ to a site $n$.

When we have the density matrix $\rho(t)$ at a time $t$ from a numerical simulation as described above, we can use Equation~(\ref{total_current}) to calculate the currents between the sites.
As described in detail in Ref.~\cite{roden2015prob_current}, the first term in Equation~(\ref{total_current}) stems from the unitary part of the dynamics -- that is, in our model of PSII from the unitary transfer driven by the electronic coupling between the pigments.
The second term stems from dephasing through the coupling to the vibrations and gives zero, that is, the dephasing does not contribute to the population currents.
The third term stems from electronic relaxation processes, that is, it describes the transfer of population to the radical pair states enabling the charge separation in the reaction center, as well as the radiative and non-radiative decay to the ground state.

We see in Equation~(\ref{total_current}) that the unitary part of the currents comes entirely from the coherence between the sites -- more precisely the imaginary part of the coherences.
The relaxation part of the currents that describe the transfer to the RP states, on the other hand, comes entirely from the populations of the sites, as can be seen in the third term in Equation~(\ref{total_current}).
This means that the energy transfer from LHCII to CP43 and to the RC derives entirely from the coherence between the sites, whereas the transfer to the RP states is entirely through the population.

When the coupling to the vibrations is strong compared to the electronic inter-pigment coupling, the resulting strong dephasing destroys a large part of the coherence between the pigments.
We see from Equation~(\ref{total_current}) that this will decrease the energy transfer between the pigments, because the unitary contribution to the population current, which in our model amounts to all the energy transfer between the pigments, decreases since it is entirely described by the coherence.
This suppression of the transport at strong coupling to the vibrations is well known and often explained in the picture of the quantum Zeno effect~\cite{rebentrost2009_033003}: 
the electronic excitation interacts with the vibrations -- is ``measured'' by the vibrations -- so strongly (i.e., at such a high rate) that the excitation is forced to stay in the original state and cannot move anymore. 

To obtain the overall net currents between the different proteins, we first calculate the currents $j_{mn}(t)$ between individual sites $m$ and $n$ of the different proteins and then sum these currents up.
That is, the current $J_{AB}(t)$ between subcomplex $A$ and a subcomplex $B$ is given by
\begin{equation}
\label{inter_protein_current}
  J_{AB}(t) = \sum_{m\in A}\sum_{n\in B} j_{mn}(t), 
\end{equation}
where when the current $J_{AB}(t)$ is positive, there is a net flow from $A$ to $B$, and when $J_{AB}(t)$ is negative, there is a net flow from $B$ to $A$.

\section{Energy transfer: simulation results}
\label{sec_simul_results}

In order to allow for a full quantum simulation of energy transfer, we construct a minimal model of light harvesting in PSII. 
In their natural environment (the thylakoid membrane), PSII supercomplexes are surrounded by light harvesting proteins (LHCII). 
Excitation energy transfer from the surrounding LHCII pigments into the supercomplex occurs via the light harvesting protein mechanically bound to the core complex (LHCII-S and LHCII-M). 
Our minimal model of the process of excitation energy transfer from the surrounding light harvesting proteins to the reactions center contains one LHCII monomer, the complete CP43 protein and a slightly truncated RC.
We use the ZOFE master equation to calculate the time-dependent density matrix in the space of the electronic states of the pigments, as described in the previous section.
This enables us to analyze the transfer of the electronic excitation energy and the electronic coherence between the pigments.
 
It is important to differentiate between the characteristics of excitation energy transfer and the functional behavior of the light harvesting apparatus. 
In low light conditions, plant growth depends on efficient light harvesting -- creating as many productive photochemical events in the RC as possible per photon absorbed~\cite{blankenship2014molecular}. 
As such, in this paper, we will equate the fraction of excitations that cause productive photochemistry (i.e. reach the RP2 state) with the function of the light harvesting antenna. 
A decrease in the fraction of excitations reaching RP2, then, is a decline in the functional behavior of the photosynthetic assembly. 
It follows that changes in the precise nature of the excitation dynamics do not necessarily result in changes to the functional behavior. 
Two different excitation dynamics can still give rise to equivalent quantities of excitation causing irreversible electron loss (reaching the RP2 state). 
As a result, to show that certain terms in an equation of motion are important to photosynthesis requires both that they change the excitation dynamics and that they increase (or decrease) the fraction of excitation that reaches RP2, the latter being a measure of how much they influence the function of photosynthesis.
In particular, we use the amount of population that reaches the second radical pair state RP2 in the reaction center within 1 ns as a measure for efficiency of the transport, since this state undergoes irreversible charge separation in our model. 
(We note, however, that fluorescence-lifetime measurements on PSII are only sensitive to the total population remaining in chlorophyll excitation. 
As a result, the experiment cannot directly differentiate the kinetics of electron transport, and the assignments to RP1 and RP2 must be considered phenomenological~\cite{bennett2013_9164}.)

In our first study, we describe excitation energy transfer when initial excitation occurs in a pigment-protein complex outside the supercomplex and subsequent energy transfer occurs into the core complex via the attached LHCII trimer. 
We therefore initiate excitation on two chlorophyll A molecules in the LHCII monomer (sites 7 and 10) to represent excitation that is entering the system from the neighboring LHCII monomers, since these sites are assumed to have a large rate of transfer with the neighboring LHCII monomers~\cite{bennett2013_9164}.
Each of the sites 7 and 10 carry half of the initial population, and we choose the initial state to not contain any coherence between the sites.
Thus, all off-diagonal elements of the initial density matrix are zero in the site basis. 
The dynamics following delocalized initial excitation will be discussed in Section~\ref{sec_deloc_init_state}.

\subsection{Population}

\begin{figure}
\centering

\hspace{-0.02\mylenunit}\includegraphics[width=0.5\mylenunit]{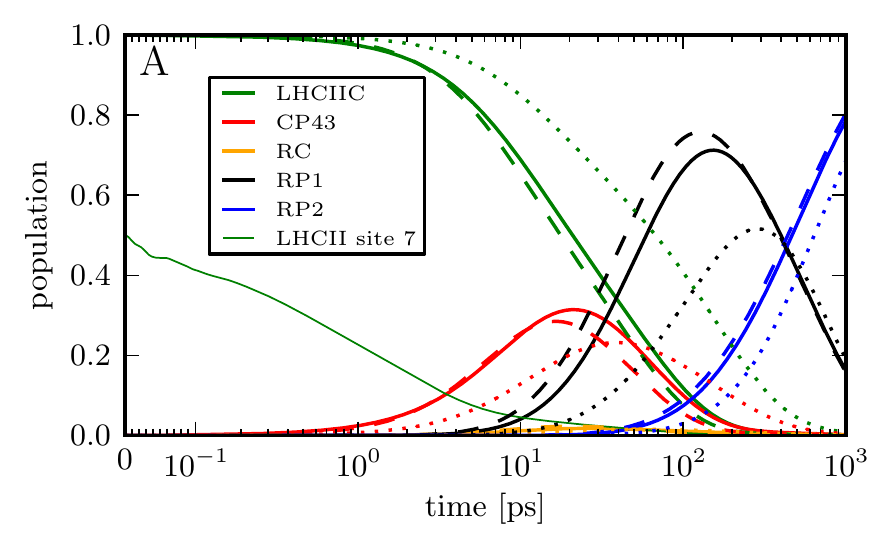}\hspace{-0.02\mylenunit}
\includegraphics[width=0.5\mylenunit]{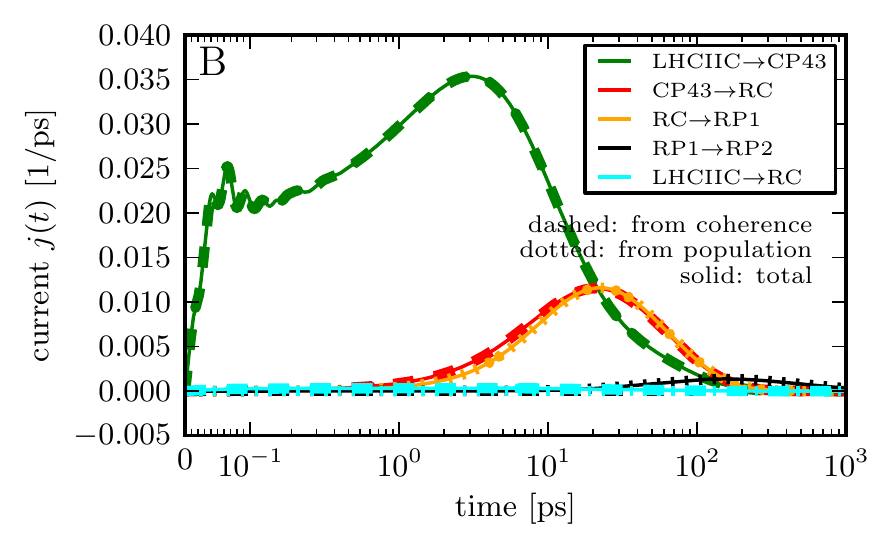}\\
\raisebox{0.05\mylenunit}{\includegraphics[width=0.5\mylenunit]{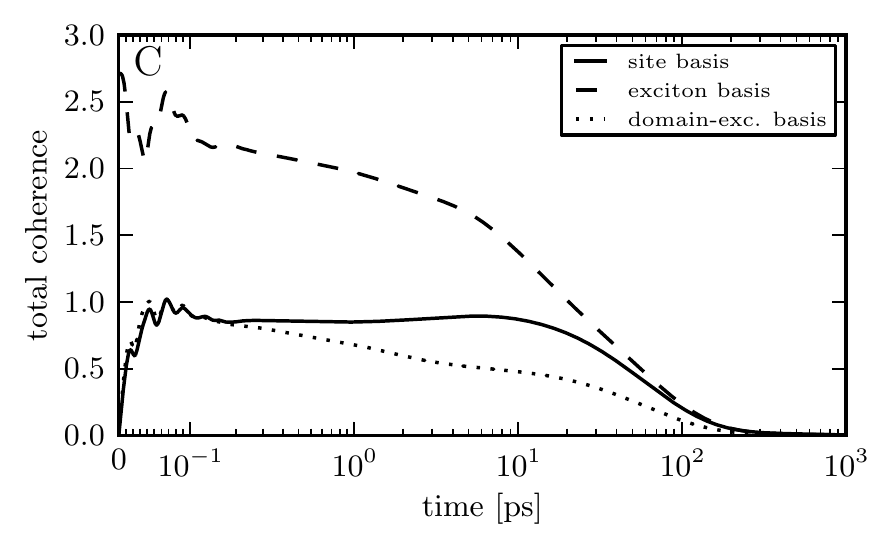}}
\includegraphics[width=0.45\mylenunit]{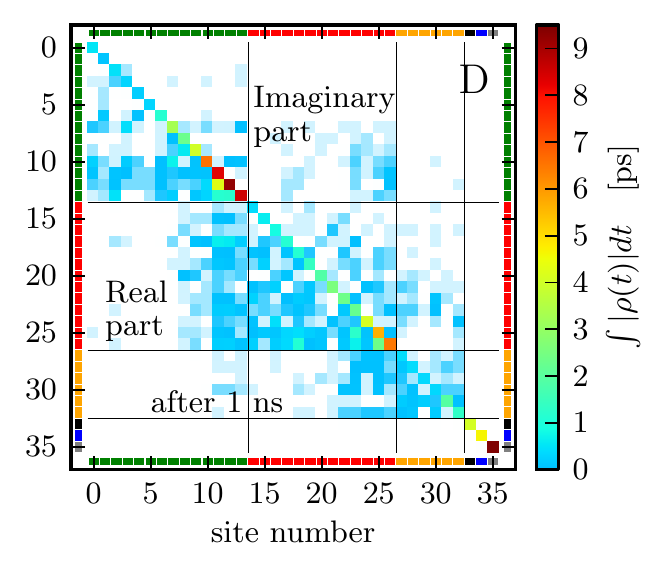}

\caption{Energy transfer dynamics calculated with the ZOFE master equation, for the electronic model and vibrational environment described in Section~\ref{sec_model}.
Initially, only sites 7 and 10 are excited, with the same amount of population on both pigments, and without initial coherence in the site basis.
{\bf Figure~A:} populations in the different proteins LHCIIC, CP43, and in the RC, and of the two radical pair (RP) states RP1 and RP2.
For comparison, the results from the generalized-F\"{o}rster-modified-Redfield calculation of Ref.~\cite{bennett2013_9164} (dashed lines) and of a pure F\"{o}rster calculation (dotted lines) are shown.
(Up to 0.1 ps, the time axis is linear; after that it is logarithmic).
{\bf Figure~B:} population currents between the proteins and RP states, calculated from the time-dependent density matrix using Equations~(\ref{total_current}) and~(\ref{inter_protein_current}).
The coherence and population contributions to the currents are shown.
(The currents have the directions specified in the legend when they are positive, and the reverse direction when they are negative).
{\bf Figure~C:} total amount of coherence $C(\rho)=\sum_{n,\,m\neq n}|\rho_{n,m}|$.
Solid line: in site basis. 
Dashed: exciton basis. 
Dotted: in the domain-exciton basis used for the F\"{o}rster-Redfield simulation of Ref.~\cite{bennett2013_9164}.
{\bf Figure~D:} elements of the density matrix in the site basis integrated over time, up to 1~ns. 
On the upper triangle, the absolute values of the imaginary parts of the coherences are shown, that is, $\int |{\rm Imag}(\rho_{nm})| dt$.
On the lower triangle, the real parts $\int |{\rm Re}(\rho_{nm})| dt$ are shown, and on the diagonal the populations $\int \rho_{nn} dt$ are shown. 
(The RP1 and RP2 populations are scaled down to 1\% of their true values.
As in Figure~\ref{fig_structure_and_ham}, the coloring at the edges indicates to which protein the sites belong, and the black lines mark the corresponding blocks of the matrix).
}
\label{fig_ZOFE_transfer}
\end{figure}

To investigate the (long-range) excitation energy transfer between the complexes and the transfer of energy to the charge separated states (RP1 and RP2), we consider the population of the electronic excitation in the individual complexes and the population of the RP states over time.
Figure~\ref{fig_ZOFE_transfer}A shows these population dynamics.
The solid lines in Figure~\ref{fig_ZOFE_transfer}A are the populations on each complex calculated with the ZOFE master equation.
The population of site 7 in LHCII is shown separately (thin green line).
Nearly all excitation (population) is transferred from LHCII through CP43 and RC to the first radical pair state RP1 within about 200 ps.
Irreversible transfer from RP1 to RP2 takes much longer because of the much weaker coupling between RP1 and RP2.
After 1 ns, 80\% of the population has been transferred to RP2.
Within this time, only a small fraction of less than 5\% is lost through radiative and non-radiative decay, in keeping with the relatively slow dynamics of excitation loss in a photosynthetic apparatus.
As we shall see in the following discussion, this separation of timescales between excitation energy reaching the RC and the relatively slow loss mechanisms of fluorescence and non-radiative decay is a key feature of efficient light harvesting in PSII.

The dashed lines in Figure~\ref{fig_ZOFE_transfer}A show the population calculated with the modified-Redfield/generalized-F\"{o}rster (MRGF) method of Ref.~\cite{bennett2013_9164}.
In the MRGF calculations, pigments are grouped together into ``domains''. 
The assignment of pigments to different domains is determined by a threshold electronic coupling and the distribution of site energies.
Within these domains, the dynamics are treated with modified Redfield theory. 
Between the domains, the transport is calculated based on transfer rates between the exciton states of one domain (electronic eigenstates of the domain) and the exciton states of another domain. 
These rates are determined using generalized F\"{o}rster theory, i.e., employing the overlaps between emission and absorption spectra of the different domains~\cite{bennett2013_9164}.

As we can see in Figure~\ref{fig_ZOFE_transfer}A, the population dynamics from the MRGF calculation -- especially the RP2 population -- agree fairly well with that of the ZOFE simulation, even though the former is a much simpler classical rate equation calculation.
This is consistent with the findings of Ref.~\cite{jesenko2013_174103} that classical rate equations are adequate for the calculation of transport efficiencies when they are properly derived from the full quantum description.

In Figure~\ref{fig_ZOFE_transfer}A, we also show the result from a pure F\"{o}rster calculation (dotted lines) for comparison, since this simple, perturbative method is often used to calculate energy transfer in biological systems.
The pure F\"{o}rster calculation is based on rates obtained from the overlaps between emission and absorption spectra of each pair of pigments (i.e.\ donor and acceptor pigments)~\cite{forster1948energy, may2008charge}.
This simple approximate method can give reasonable results in a regime where the coupling between donor and acceptor pigments is weak compared to the coupling to the vibrations.
However, since this condition is not met for many pigment pairs in PSII, especially inside the more strongly coupled domains, we expect the pure F\"{o}rster calculation to be less accurate than the ZOFE calculation as well as the MRGF calculation of Ref.~\cite{bennett2013_9164}. 

In Figure~\ref{fig_ZOFE_transfer}A we do indeed see a significant deviation of the pure F\"{o}rster calculation (dotted lines) from the other simulations.
Furthermore, the transport efficiency (population reaching RP2 within 1~ns) calculated with pure F\"{o}rster is lower.

The observation that the MRGF and ZOFE calculations are in close agreement with each other, while the pure F\"{o}rster calculation gives a lower transport efficiency, supports the key role of exciton delocalization in excitation transport through PSII.
The dynamic effect is sometimes referred to as supertransfer, since it occurs when strongly coupled pigments allow an excitation to coherently delocalize within the donor domain, thereby enhancing the rate of transport by up to a factor equal to the number of pigments in the donor domain (see e.g.\ Ref.~\cite{kassal2013_362} and references therein).
This enhancing effect of supertransfer due to excitonic delocalization within a subcomplex or domain is naturally included in the MRGF and ZOFE calculations, but it is not taken into account in the pure F\"{o}rster description, which thus leads to a lower transport efficiency.

\subsection{Population currents}

Figure~\ref{fig_ZOFE_transfer}B shows the population currents $j(t)$ between the proteins and from the reaction center to the radical pair states.
These currents are calculated from the time-dependent density matrix using Equations~(\ref{total_current}) and~(\ref{inter_protein_current}).

From Equation~(\ref{total_current}), we know that the energy transfer from LHCII to CP43 and on to the RC is entirely through the coherence between the pigments, and the transfer to the RP states is entirely through the population.
This is combined by Figure~\ref{fig_ZOFE_transfer}B, where the solid lines are the total currents, and the dashed and dotted lines show the coherence and population contribution, respectively.

We see that all currents between complexes are positive over the entire time range.
That means that the net flow of energy is directed towards the RP2 state at all times and no temporary intercomplex backflow occurs.
Because of the high rate of transport between the RC and the RP1 state, in Figure~\ref{fig_ZOFE_transfer}B the current from RC to RP1 almost completely follows the current from CP43 to the RC; there is only a very small delay of the current to RP1.
That means the population is almost passed right through the RC to RP1.
The current to the RP2 state has a smaller magnitude but stretches over a longer time range.
Therefore the overall population that is transferred to RP2 during this time span, which is simply the current integrated over time, is not much smaller than the population transferred to the RC, namely 80\% versus~$\gtrsim 95\%$. 
As expected, the {\it direct} current between LHCII and the RC (cyan curve in Figure~\ref{fig_ZOFE_transfer}B) is zero, due to the large distance between the LHCII and RC pigments. 

\begin{figure}
\centering
\includegraphics[width=1.1\mylenunit]{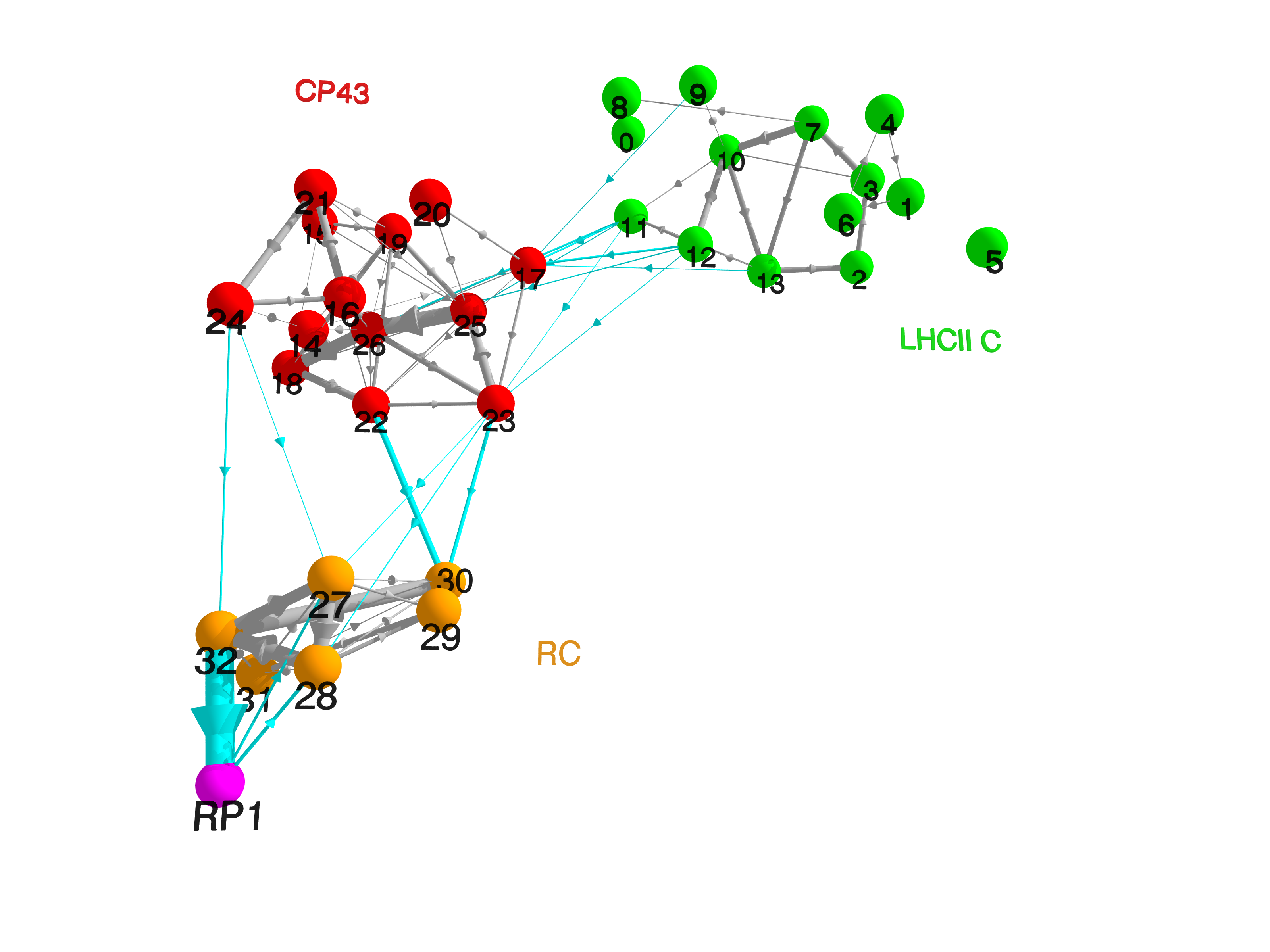}

\caption{Population currents between the individual chromophores, integrated over time, that is, $\int j_{mn}(t)\,dt$.
Integrated up to a time of 1~ns, and for the same simulation as Figure~\ref{fig_ZOFE_transfer}.
As in Figure~\ref{fig_ZOFE_transfer}, initially only pigments~7 and~10 are excited.
The arrows show the directions and their thickness the relative magnitude of the integrated currents.
Only integrated currents larger 0.03 are shown.
The gray arrows show the currents inside the proteins and the cyan arrows the ones between the proteins.
The relative positions of the pigments in 3D are obtained from Ref.~\cite{bennett2013_9164}.
(The labeling of the pigments is the same as in Fig.~\ref{fig_structure_and_ham}.)
}
\label{fig_ZOFE_transfer_currents}
\end{figure}

To see the individual pathways of the energy transfer, we integrate the population currents between the individual pigments over time. 
The result is shown in Figure~\ref{fig_ZOFE_transfer_currents}.
We see in Figure~\ref{fig_ZOFE_transfer_currents} that there is a complex network of pathways rather than just one or two dominant pathways.
Previous simulations using the MRGF model have shown that transport through PSII does not occur by a single, obligate pathway~\cite{bennett2013_9164}.
Our current simulations demonstrate that there are multiple transport pathways from LHCII to CP43, and from CP43 to the RC.
However, particularly within complexes, there are a few pathways that transport a particularly large amount of excitation energy, identified in Figure~\ref{fig_ZOFE_transfer_currents}.

Remarkably, in Figure~\ref{fig_ZOFE_transfer_currents} we see that there are a number of relatively strong currents that run in a circular fashion, where energy is transported in a loop between pigments. 
For instance, there is a loop in LHCII from pigment~7 to 13 to 2 to 3, and back to 7.
In CP43, there are several such loops. 
For instance, from 25 to 26 to 23, and back to 25.
And there is a longer one from 25 to 26 to 18 to 22 to 23, and back to 25.
There is also one from 16 to 21 to 24, and back to 16.

In the reaction center, there is a strong circular current from 32 to 27 to 28, and back to 32.
We also see in Figure~\ref{fig_ZOFE_transfer_currents} that in the reaction center overall the dominant net flow to the RP1 state is via pigment 30 and 32. 
Even though pigments 27 and 28 are part of the strong circular current, they do not appear to contribute significantly to the net transport to RP1.
There is even a net backflow from the RP1 state to the pigments 27 and 28, which acts to decrease the efficiency of the transport.
Pigments 27 and 28 receive a small amount of population from CP43 pigments, but the backflow from RP1 to these two pigments is larger and therefore seems to outweigh their contribution to the transport to RP1.
In the light of these pathways, it appears that the efficient excitation energy transfer to and in the RC relies mainly on the two pigments 30 and 32, and that therefore other reaction center pigments may rather be needed for their role in the electron transport (described by the RP states in our model) than for the excitation transfer.
To test this hypothesis, we ran additional simulations where -- except for the pigments 30 and 32 -- we decoupled all RC pigments from the rest of the pigments, inhibiting electronic excitation transfer to all RC pigments but 30 and 32, while leaving the coupling to the RP states unmodified.
We found the resulting efficiency of the energy transport to the RP2 state in this restricted model to be indeed the same as in the original model where the excitation transport to the other RC pigments was allowed. 
This shows that only the two pigments 30 and 32 may be needed for the excitation energy transport to the RP states, and the role of other RC pigments appears to be mainly in the electron transport (described by the RP states).

\subsection{Amount of coherence}

According to the population current description of Equation~(\ref{total_current}), the {\it imaginary component} of coherence between pigments is responsible for the population currents through PSII.
In the ongoing discussion about the role of coherence in the energy transfer in biological systems, coherence and its contribution to the energy transfer has so far been quantified according to different metrics that generally do not look separately at the imaginary and real parts of the coherence (see e.g.\ Refs.~\cite{baumgratz2013quantifying, kassal2013_362} for quantification and discussion of coherence).
However, as described in detail in Ref.~\cite{roden2015prob_current}, the imaginary and the real part of the coherence have very different effects -- the imaginary part drives the transfer, whereas the real part is related to the transfer-diminishing localization effects in the presence of energy gaps between the pigments.  
Therefore, we shall not only quantify the total coherence, but also look separately at its imaginary and real parts.

Solving the ZOFE master equation for the full quantum dynamics gives the time-dependent electronic density matrix and therefore allows direct quantification of the coherences, the off-diagonal elements of the density matrix in the respective basis of interest.
We first quantify the total amount of coherence by the sum 
\begin{equation}
\label{sum_coher}
  C(\rho) = \sum_{n, m\neq n}|\rho_{nm}|
\end{equation}
over the absolute values of all off-diagonal elements of the density matrix $\rho$, which provides an intuitive measure for the overall amount of coherence.
This ``$l_1$-norm of coherence'' has been widely used in previous studies and is reviewed in Ref.~\cite{baumgratz2013quantifying}.

Aside from the coherence in the site and exciton bases, we are also interested in the coherence in the domain-excition basis of Ref.~\cite{bennett2013_9164}.
This is of interest since in the MRGF calculation of Ref.~\cite{bennett2013_9164}, a classical rate equation is used to describe long-range intercomplex transfer that explicitly contains only the populations in the domain-exciton basis, but not the coherences.
Thus, coherence in this basis is not explicitly taken into account in the calculation giving the population dynamics in Figure~\ref{fig_ZOFE_transfer}A (dashed lines), nevertheless good agreement with the ZOFE population dynamics is achieved.
This is consistent with Ref.~\cite{jesenko2013_174103}, where it is discussed that the overall population dynamics, or at least the energy transfer efficiency, that is, the population transferred to the radical pair states, should be retained under a proper transition from the full quantum description to the classical rate description where coherence is not explicitly taken into account.

In Figure~\ref{fig_ZOFE_transfer}C, we show the total amount of coherence measured by Equaiton~(\ref{sum_coher}) for the site basis, the exciton basis, and the domain-excition basis.
The domain-exciton coherence shows the coherence not (explicitly) taken into account in the MRGF calculations of Ref.~\cite{bennett2013_9164}.

Since the initial state is an incoherent state in the site basis, at time zero there is no coherence in the site basis.
Similarly, in the domain-exciton basis, the coherence is initially zero.
(Sites 7 and 10, which carry the initial excitation, belong to the same domain that consists of only these two sites.) 

In the exciton basis, on the other hand, there is initial coherence at time zero, because the excitation localized on sites 7 and 10 is expressed through a coherent superposition of exciton states that each have part of their excitation on sites 7 and 10, but are also delocalized over other sites, mainly sites of the LHCII antenna protein. 

In both site and domain-exciton basis, the amount of coherence rises initially, because the dipole-dipole interaction between the pigments drives the initially localized state towards a more delocalized state.
During the transport dynamics, the total electronic coherence decreases due to interaction with the vibrations, and later due to the accumulation of population in the radical pair states.

Next, we look at the time-integrated coherences between the individual pigments and also separately at the imaginary and real parts of the coherence, because of their different role in the energy transfer.
To this end, we time-integrate the absolute value of the imaginary and real parts of the elements of the density matrix in the site basis during the first 1 ns of its propagation (at which point, as we have shown above, the majority of the excitation has been irreversibly trapped at RP2).
In Figure~\ref{fig_ZOFE_transfer}D, the resulting site-basis elements of the time-integrated density matrix are shown as a color matrix.
The time-integrated pigment populations are the diagonal elements. 
One can see that there is an accumulation -- a ``damming-up'' -- of population on the energetically lowest sites of each protein.
Such a damming-up effect occurs because the transport of excitation that is fast inside the complexes, where the couplings between the pigments are relatively strong, is slowed down between the complexes.
(It also depends on the coupling to the vibrations and also on the gap between the ``donor'' protein's pigment energies and the ``acceptor'' protein's pigment energies.)
The upper triangle in Figure~\ref{fig_ZOFE_transfer}D shows the (absolute values of) the imaginary parts of the coherences between the sites integrated over time.
That is, it shows $\int |{\rm Im}(\rho_{n,m\neq n})|\, dt$.
As seen in Equation~(\ref{total_current}), the imaginary parts of the coherences give the unitary contribution to the population currents between the sites.
That is, the imaginary parts of the coherence constitute the whole of the energy transfer, except for the transfer from the reaction center to the radical pair states, which is decribed by relaxation.

The lower triangle in Figure~\ref{fig_ZOFE_transfer}D shows the real parts of the coherence, that is, $\int |{\rm Re}(\rho_{n,m\neq n})|\, dt$.
We see that between many sites, the real part of the coherence is larger than the imaginary part.
As discussed in Ref.~\cite{roden2015prob_current}, the real part of the coherence is related to localization effects in the energy transfer due to the energy gaps between the pigments. 
The smaller the real part of the coherence, the smaller are these localization effects that can hinder the energy transfer.
(Elements that are very small are ignored in the matrix in Figure~\ref{fig_ZOFE_transfer}D).

There are relatively strong imaginary parts of coherences visible in the blocks that show the coherence between LHCII sites and CP43 sites, and also between CP43 sites and RC sites.
These are the coherences that are responsible for the currents between these complexes, shown by the green and red curves in Figure~\ref{fig_ZOFE_transfer}B.
In Figure~\ref{fig_ZOFE_transfer}D, we see that there are even non-zero coherences between LHCII sites and reaction center sites. 
However, these are very weak and the resulting net population current that they cause between LHCII and the RC appears to be close to zero in Figure~\ref{fig_ZOFE_transfer}B.

\section{Robustness of energy transport and function in PSII}
\label{sec_robustness}

In the previous section we have described excitation energy transfer when initial excitation occurs in a pigment-protein complex on the periphery of the supercomplex and subsequent energy transfer occurs into the core complex via the attached LHCII trimer. 
In this Section, we shall explore how robust the excitation energy dynamics and the functional behavior of PSII are to changes in the initial state and the electron-phonon coupling.

\subsection{Delocalized initial excitation}
\label{sec_deloc_init_state}

There is an ongoing discussion about the initial conditions of light absorption in light-harvesting systems, and how they affect the subsequent transport dynamics~\cite{kassal2013_362, jesenko2013_174103, han2013_8199}.
In particular, initial conditions in nature -- excitation by sunlight -- versus initial conditions in laser spectroscopy experiments -- excitation through coherent laser light -- have been discussed by a number of authors~\cite{kassal2013_362, jesenko2013_174103, han2013_8199}. 

Motivated by this, we now investigate the influence of the initial state on the long-range energy transfer in PSII.
So far, we have considered initially localized excitation of only two pigments in the LHCII, with initial coherence in the exciton basis but not in the site basis.
Now, we contrast this by considering an initial state that is a completely delocalized superposition of all exciton states of all three proteins.
We choose this initial state to be an (equally weighted) {\it incoherent} superposition of all exciton states -- i.e., no initial coherence in the exciton basis.
Thus, we take an initial density matrix, where in the exciton basis all diagonal elements (populations) are $1/N_{\rm sites}$ ($N_{\rm sites}$ is the number of sites), except for the population of the radical pair states and the ground state, which are zero, and with all off-diagonal elements (coherences) set to zero.

\begin{figure}
\centering
\hspace{-0.02\mylenunit}\includegraphics[width=0.5\mylenunit]{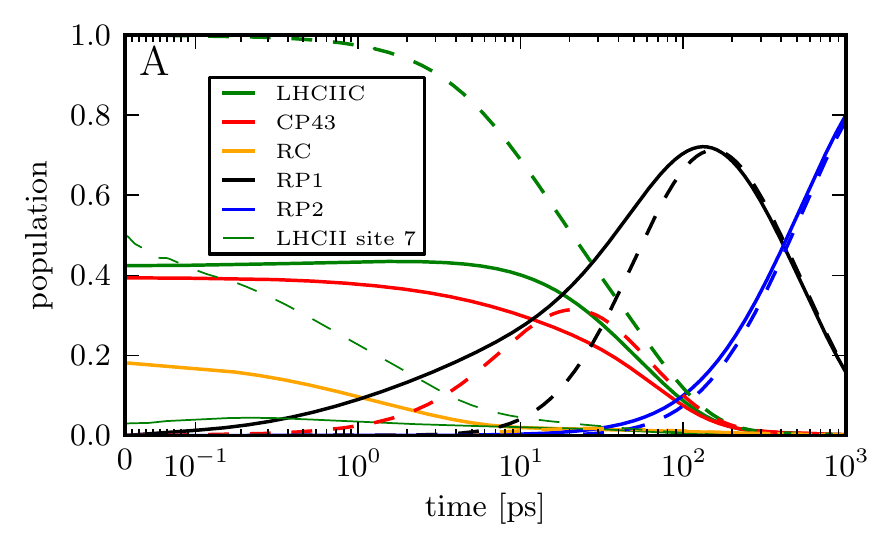}\hspace{-0.02\mylenunit}
\includegraphics[width=0.5\mylenunit]{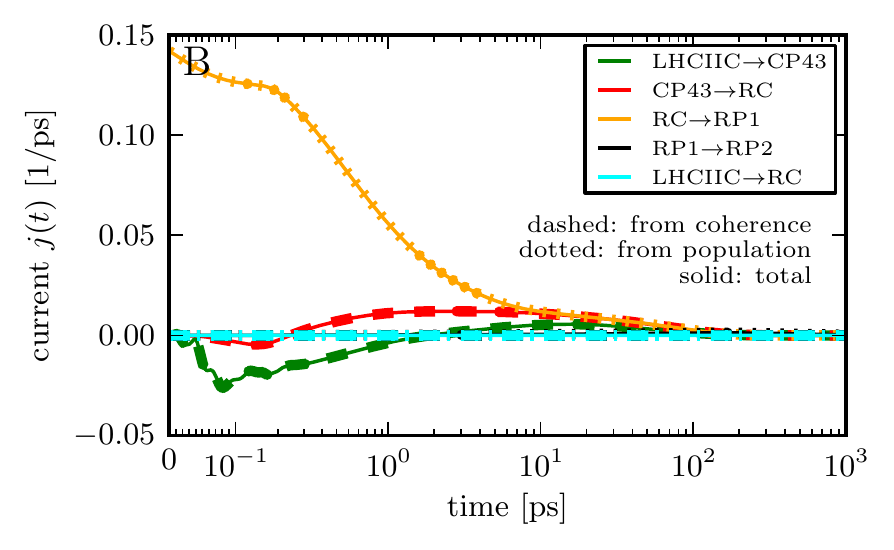}\\
\raisebox{0.05\mylenunit}{\includegraphics[width=0.5\mylenunit]{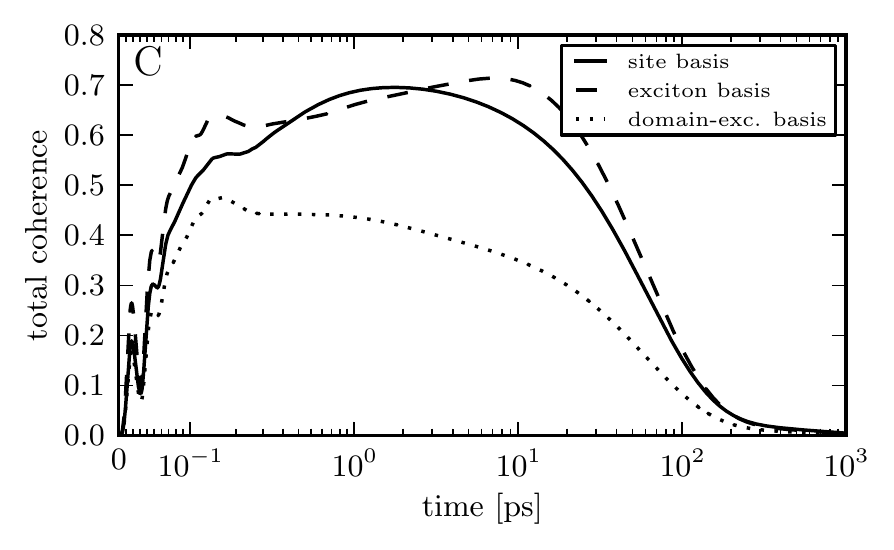}}
\includegraphics[width=0.45\mylenunit]{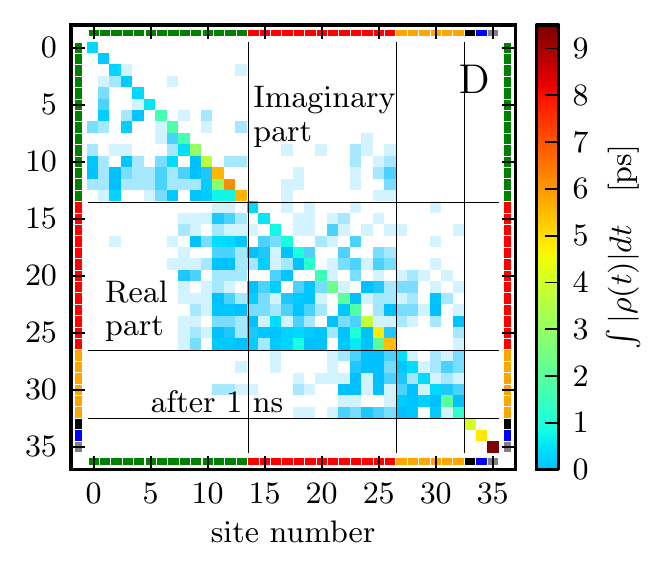}

\caption{As Figure~\ref{fig_ZOFE_transfer}, but the initial state is an (equally weighted) incoherent superposition of all exciton states of all three proteins, LHCII, CP43, and the RC.
For comparison, the dashed lines in Figure~A show the ZOFE dynamics from Figure~\ref{fig_ZOFE_transfer}A for the localized initial state.}
\label{fig_ZOFE_transfer_initAllExcIncoher}

\end{figure}

Since the exciton states are the eigenstates of the system Hamiltonian, this state would persist in the course of purely unitary system time propagation.
But the coupling to the vibrations, the coupling to the radical pair states, and the radiative and non-radiative decay to the ground state cause the initial state to evolve in a non-unitary manner, driving the excitation transport through PSII.
These dynamics lead to the creation of excitonic coherence in the course of time evolution, even though it is not present initially.
The transport dynamics that result from this incoherent, delocalized initial state are summarized in Figure~\ref{fig_ZOFE_transfer_initAllExcIncoher}.

The population dynamics are shown in Figure~\ref{fig_ZOFE_transfer_initAllExcIncoher}A and are compared to the dynamics resulting from the localized initial state (Figure~\ref{fig_ZOFE_transfer}A). 
We see that for the delocalized initial state, the population dynamics of the pigments now look quite different.
In particular, 40\% of the initial excitation is in the CP43 protein and almost 20\% is in the reaction center pigments. 
As a result, 60\% excitation is spatially closer to the reaction center than in the previous simulation. 
This results in a much earlier rise in the RP1 population (black line) than seen in our previous simulation (black, dashed line). 
Some of this excitation, however, is initiated in higher energy states allowing for back-transfer from CP43 to LHCII as can be seen in Figure~\ref{fig_ZOFE_transfer_initAllExcIncoher}B. 
The green curve, representing the flow of excitation from LHCII to CP43 is negative now for an initial period of time, showing that there is a flow of excitation from CP43 back to LHCII. 
That is to say, after initial excitation there is a net excitation transport in the opposite direction as seen before for the localized initial state, namely away from the reaction center.

\begin{figure}
\centering
\includegraphics[width=1.1\mylenunit]{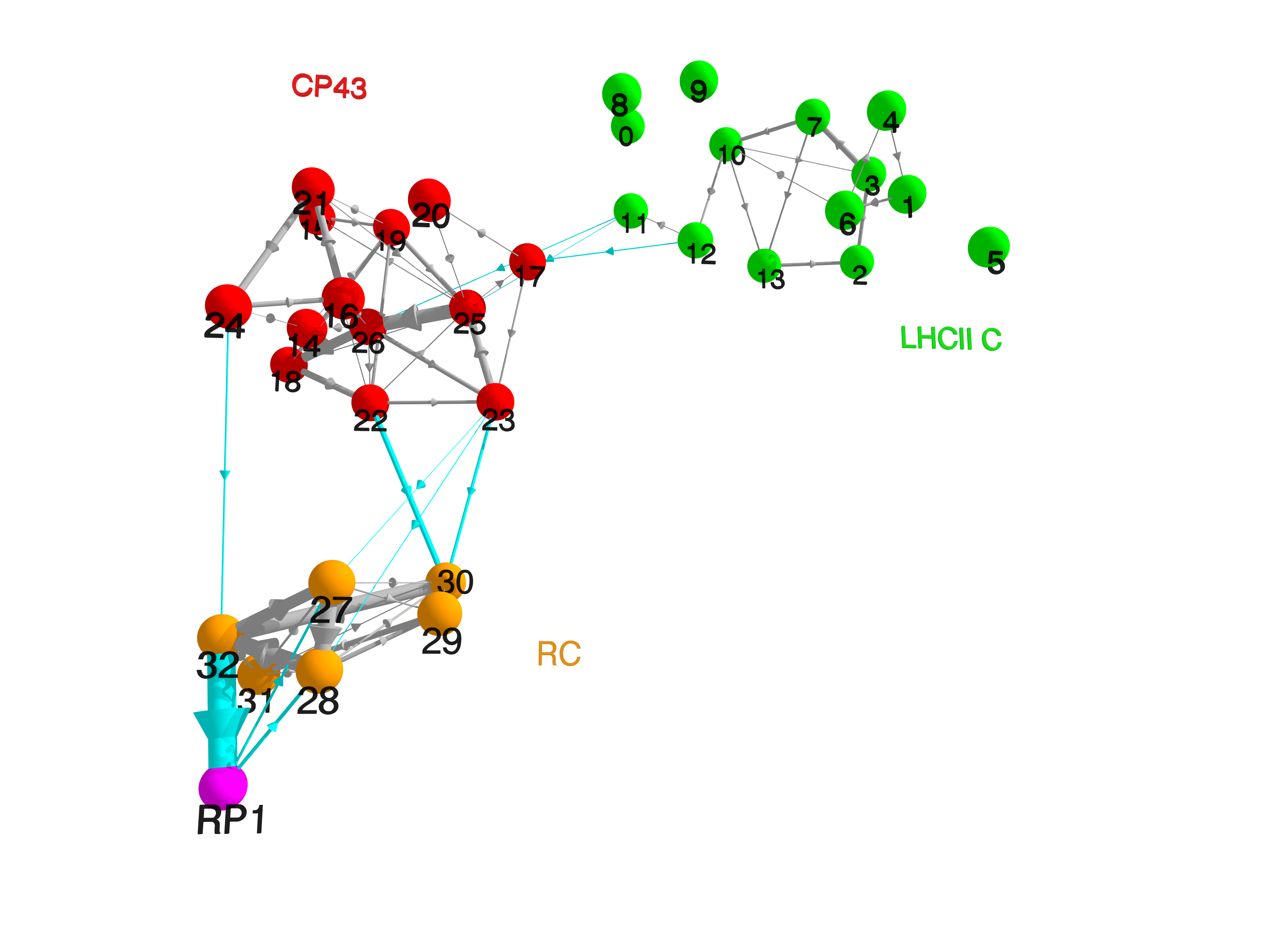}

\caption{As Figure~\ref{fig_ZOFE_transfer_currents}, but the initial state is an (equally weighted) incoherent superposition of all exciton states of all three proteins, LHCII, CP43, and the RC.
}
\label{fig_ZOFE_transfer_currents_initAllExcIncoher}
\end{figure}

Figure~\ref{fig_ZOFE_transfer_currents_initAllExcIncoher} shows the time-integrated population currents between the individual pigments for the delocalized initial state, and should be compared to Figure~\ref{fig_ZOFE_transfer_currents} for the localized initial state.
The currents inside LHCII and from LHCII to CP43 are smaller now than before. 
This is expected, since now only part of the initial population is in LHCII. 
Aside from these changes, however, the overall pattern of the pathways is very similar to what we have seen for the localized initial state before. 
The currents have the same directions as before, and even their relative magnitudes seem very similar. 
This is a remarkable result, since we know in this initial condition, every pigment has an equal initial excitation. 
Nevertheless, both the short-range (intra-complex) and long-range (inter-complex) energy transfer are very similar for both initial conditions. 
Interestingly, this suggests that while there is not a single obligate pathway of excitation energy transport within PSII -- there is a select subset of PSII pigments that are used to create an excitation transport network. 
This observation may allow future work to differentiate between pigments that are involved in long-range transport from those that are predominately only associated with absorbing light and then transferring excitation locally (and hence showing little to no net flux in either of our initial conditions). 

Despite the changes in excitation transport dynamics, the functional behavior of PSII is invariant to the change in initial excitation. 
In particular, the overall transport of population to the radical pair state RP2 is almost identical for both initial excitation conditions, as can be seen in Figure~\ref{fig_ZOFE_transfer_initAllExcIncoher}A.
(We note that because of the logarithmic time axis, the earlier deviation between the RP1 populations for the different initial states appears to be much larger than the deviation between the RP2 populations at later times, but in fact these deviations are of comparable magnitude.)
This shows that changing the initial state from the previously considered localized state to a delocalized state does not change the efficiency of the energy transport.
The stability of the functional behavior with respect to changes in excitation energy dynamics can be explained by the separation of timescales between the excitation energy transport and the loss mechanisms. 
Excitation is lost in PSII on a timescale of $\sim 2$~ns through a combination of fluorescence and non-radiative decay. 
Excitation transport, however, occurs on a $\sim 100$~ps timescale. 
If we approximate the excitation transport to the radical pair states by a single rate constant, then the overall behavior is described by a two-rate (three-state) model -- one rate for excitation to drive charge separation and a second rate of excitation undergoing non-radiative or radiative decay. 
Within this simplified model, the fraction of excitation that performs charge separation (as opposed to non-radiative or radiative decay) as time goes to infinity is given by
\begin{equation}
  \label{long_time_efficiency}
  P_{\rm RP2}(t\rightarrow\infty) = \frac{k_{{\rm exc}\rightarrow{\rm RP}}}{k_{{\rm exc}\rightarrow{\rm RP}} + k_{\rm decay}}
\end{equation}
-- the population reaching the RP2 state in the long-time limit -- where $k_{{\rm exc}\rightarrow{\rm RP}}$ is the rate of population transfer to the RP states and $k_{\rm decay}$ is the rate of loss through non-radiative and radiative decay.  
We will use the phrase long-time efficiency to refer to this quantity $P_{\rm RP2}(t\rightarrow\infty)$ and differentiate it from our 1~ns measure of efficiency, i.e.\ $P_{\rm RP2}(t=1\ {\rm ns})$, used in the remainder of the paper. 
In Figure~\ref{fig_ZOFE_transfer_initAllExcIncoher}A we see that at 100~ps the total excitation left on the pigments amounts to a population of $P_{\rm pig}(t=0.1\ {\rm ns})\approx 0.2$.
Inserting this value in $k_{{\rm exc}\rightarrow{\rm RP}} = -\frac{1}{t}{\rm ln}(P_{\rm pig}(t)\,e^{k_{\rm decay}t})$, we estimate the rate of excitation transport from the LHCII initial state (dashed lines) to the RP states as $k_{{\rm exc}\rightarrow{\rm RP}}\approx 15\ {\rm ns}^{-1}$, which, using Eq.~(\ref{long_time_efficiency}), gives a long-time efficiency of 97\%. 
To substantially change this value, the rate of excitation transport to the RP states would need to slow enough to become comparable to the $0.5\ {\rm ns}^{-1}$ rate of excitation loss. 
For example, slowing the rate of transport to the RP states by a factor of 2 only decreases the long-time efficiency to 93\%. 
As a result, small modulations in the rate of transport to the RP states have only a minimal influence on the long-time efficiency within our model, and thus will not substantially shift the functional behavior of PSII.

We now analyze the dependence of the amount of coherence on the initial state.
In Figure~\ref{fig_ZOFE_transfer_initAllExcIncoher}C, we see that for the delocalized initial state there is a significantly smaller amount of excitonic coherence involved in the transfer dynamics compared to the amount for the localized state in Figure~\ref{fig_ZOFE_transfer}C (note the different range of the y axis of Figure~\ref{fig_ZOFE_transfer_initAllExcIncoher}C compared to Figure~\ref{fig_ZOFE_transfer}C).
This is mainly due to the fact that there is no initial excitonic coherence present in the case of the delocalized initial state.
In Figure~\ref{fig_ZOFE_transfer_initAllExcIncoher}D, the pattern of the time-integrated coherence between the individual sites appears overall similar to that for the localized initial state in Figure~\ref{fig_ZOFE_transfer}D.
However, looking closer, there are differences, which are reflected in the large difference between the currents resulting from the delocalized and localized initial states.
We note also, that since in Figure~\ref{fig_ZOFE_transfer_initAllExcIncoher}C and~D absolute values are summed up, changes of the signs of the individual terms that would correspond to opposite directions of the corresponding population currents are not visible.
For example currents that lead to a flow of excitation back to LHCII in Figure~\ref{fig_ZOFE_transfer_initAllExcIncoher}B, which have negative signs, cannot be identified in Figure~\ref{fig_ZOFE_transfer_initAllExcIncoher}D, because the absolute value is taken before integrating the coherences. 
But we see in Figure~\ref{fig_ZOFE_transfer_initAllExcIncoher}D that just as the current between LHCII and CP43 is smaller for the delocalized initial state, so the imaginary parts of the coherences between LHCII and CP43 pigments are also weaker than before in Figure~\ref{fig_ZOFE_transfer}D.

We have also performed simulations for other initial states, including localized, delocalized superpositions with coherence in the site or exciton basis, and for localized initial excitations with specific phase relations that were previously found to lead to directed transport for models of chains of coupled sites~\cite{eisfeld2011phase}.
In all these cases, we do not find any significant change in the overall efficiency of the transport to the RP2 state, showing a remarkable robustness of the overall long-range energy transport with respect to the initial conditions.

\subsection{Influence of the coupling to vibrations}

We now investigate the influence of the coupling between electronic and vibrational degrees of freedom on the energy transfer.
The coupling to the vibrations is treated approximately in our model so that it is important to consider how sensitive the energy transfer is with respect to changes in this part of the model.
Furthermore, we expect the coupling to the vibrations to be an important component of the energy transfer dynamics, since it can enhance the transport, as is well known and has been studied for several smaller scale (single complex) systems (see e.g.\ Refs.~\cite{rebentrost2009_033003, kolli2012vibrations, chin2012coherence}).

In the following, we investigate these relationships by varying the strength of the coupling to the vibrations, that is, the magnitude of the environment spectral densities of the pigments Eq.~(\ref{spec_dens}).
For all pigments, we simply multiply their spectral densities $J_n(\omega)$ by a global factor that is the same for all of them, i.e.,
\begin{equation}
\label{factor_spec_dens}
  J_n'(\omega) = c \ J_n(\omega),
\end{equation}
where the constant factor $c$ is the same for all pigments $n$ in all proteins.
As the initial state, we take again the state where the excitation is localized on only the two pigments~7 and~10 in LHCII that we have used in Section~\ref{sec_simul_results}.

\begin{figure}
\centering
\includegraphics[width=0.54\mylenunit]{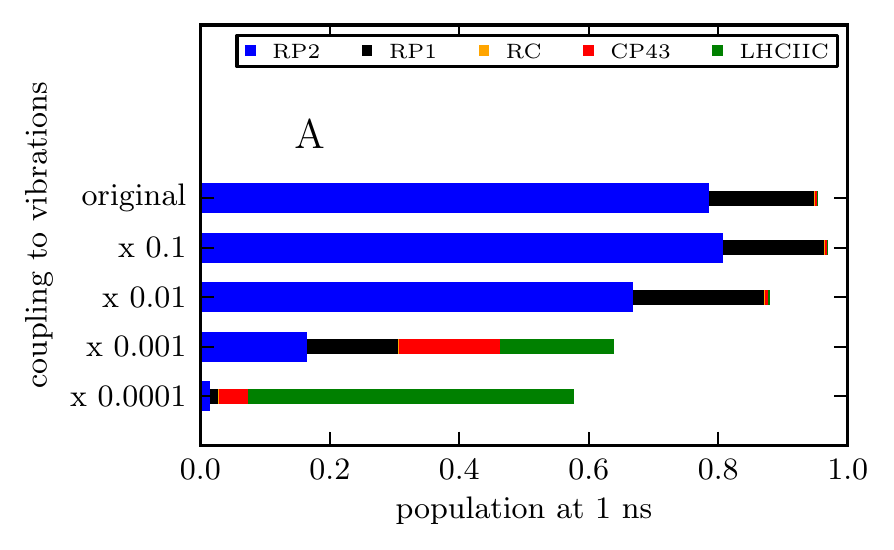}\\
\hspace{-0.02\mylenunit}\includegraphics[width=0.5\mylenunit]{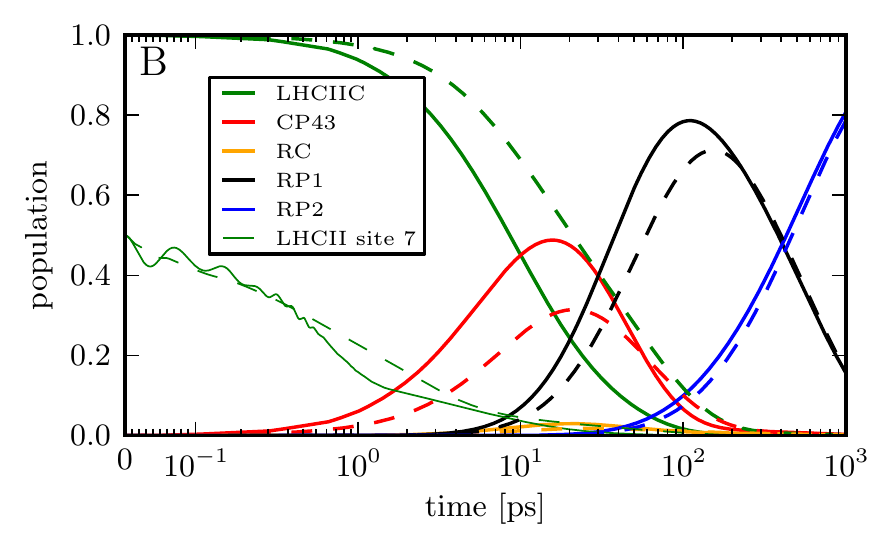}\hspace{-0.02\mylenunit}
\includegraphics[width=0.5\mylenunit]{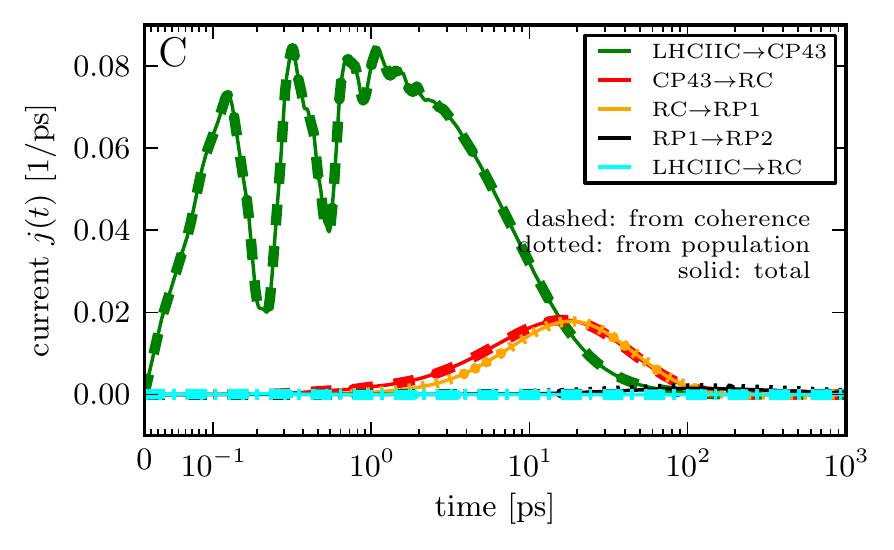}\\
\hspace{-0.02\mylenunit}\includegraphics[width=0.5\mylenunit]{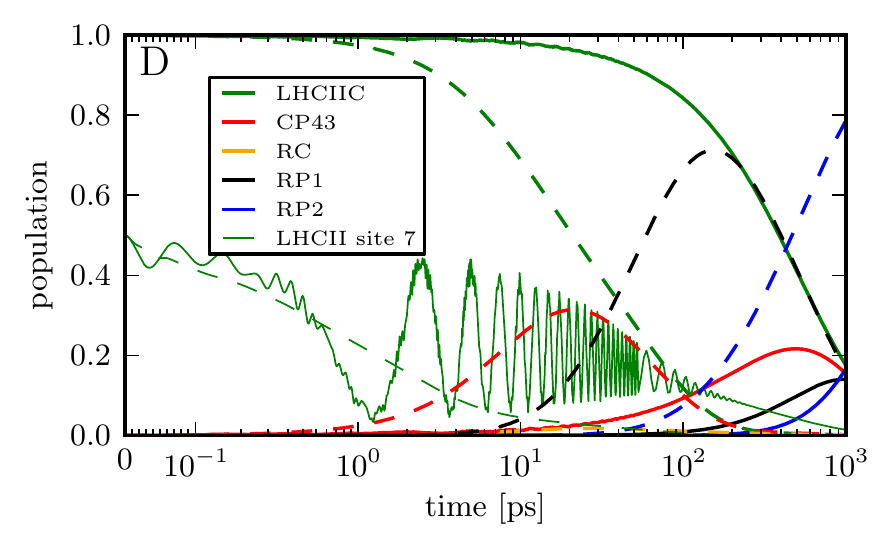}\hspace{-0.02\mylenunit}
\includegraphics[width=0.5\mylenunit]{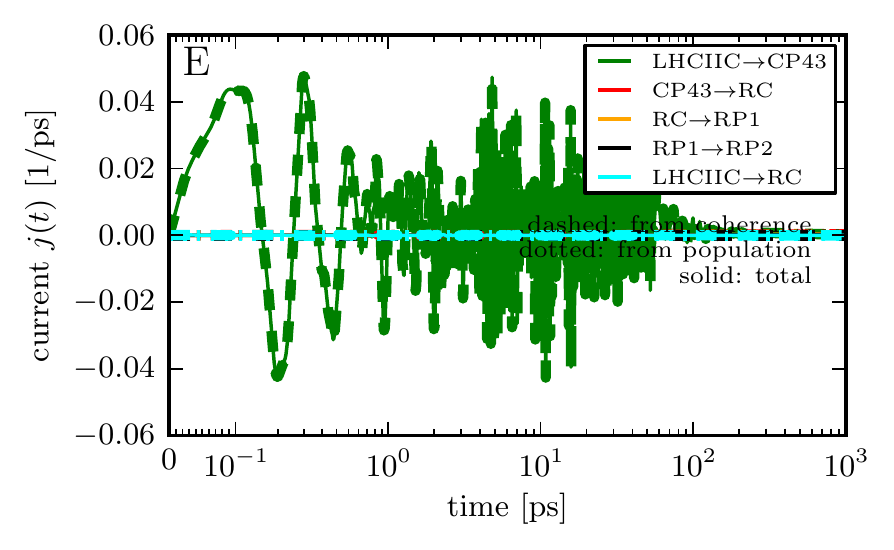}

\caption{Effect of decreasing the coupling strength to the vibrational environment by factors of ten; ZOFE simulation results.
{\bf Figure~A:} Populations at 1~ns in the different complexes and of the radical pair states RP1 and RP2.
The bars for RP2 are thicker for better distinguishability.
{\bf Figure~B and~C:} As Figures~\ref{fig_ZOFE_transfer}A, B, but the coupling to the vibrational environment is decreased {\bf by a factor of 10} compared to Figure~\ref{fig_ZOFE_transfer} (corresponding to the second bar in A).
For comparison, the dashed lines in Figure~B show the ZOFE dynamics from Figure~\ref{fig_ZOFE_transfer}A for the original coupling to the vibrations.
{\bf Figure~D and~E:} As B, C, but the coupling to the vibrational environment is decreased {\bf by a factor of 1,000} compared to the original coupling strength (corresponding to the fourth bar in A).
In D, the dynamics for the original coupling strength are again shown as dashed lines.
}
\label{fig_vary_coupl_strength_vibr_ZOFE}

\end{figure}

To investigate the effect of decreasing coupling to the vibrations on the energy transport, we decrease the coupling in stages.
Figure~\ref{fig_vary_coupl_strength_vibr_ZOFE} shows how the energy transfer changes when we decrease the coupling to the vibrations by factors of ten, i.e., $c=0.1, 0.01, 0.001, 0.0001$.
For each of these cases, Figure~\ref{fig_vary_coupl_strength_vibr_ZOFE}A shows the population of the RP2 state (blue, thick bar) and RP1 state and the total population in each of the complexes LHCII, CP43, and the RC (colored, thin bars) at a time of 1~ns.
We see that when we decrease the vibrational coupling, the transport efficiency -- the population of the RP2 state at 1~ns -- for a decrease of the coupling by factor~0.1 first increases slightly.
Then, as we further decrease the coupling to the vibrations, the efficiency of the transport decreases monotonically.
We can also see that when the vibrational coupling is smaller and smaller, there is more and more population trapped in LHCII.
It is remarkable that even when we decrease the coupling by a factor of hundred (i.e.\ $c=0.01$), the amount of population transported to RP2 within 1~ns is still higher than 60\% (compared to about 80\% in the case of original coupling strength), showing that the energy transport in PSII appears to be very robust with respect to changes of the coupling to the vibrations.
Only when we decrease the coupling further do we find that the energy transfer efficiency decreases significantly.
In the extreme case where we decrease the coupling to the vibrations by a factor of thousand (i.e.\ $c=0.001$), it goes down to less than 20\% population on RP2 after 1~ns. 
Decreasing the coupling further ($c=0.0001$), causes most of the population to be trapped in LHCII and only a very small amount of population ($\sim 2\%$) reaches RP2.
We can see in Figure~\ref{fig_vary_coupl_strength_vibr_ZOFE}A that the total length of the bars, i.e.\ the total population of the RP states and of the complexes together at 1~ns, is less than 1 (100\%).
The difference between the total length of the bars and 100\% is the amount of population that is lost due to radiative and non-radiative decay to the ground state during the time of 1~ns.
We see that this loss of population increases as the coupling to the vibrations is decreased, because the less population reaches the RP states, the more population remains in the excited electronic states of the pigments where it is subject to the decay.

For the two cases of $c=0.1$ and $c=0.001$ (the second and fourth bars from the top in Figure~\ref{fig_vary_coupl_strength_vibr_ZOFE}A), we take a closer look at the energy transport dynamics in Figures~\ref{fig_vary_coupl_strength_vibr_ZOFE}B, C (for $c=0.1$) and Figures~\ref{fig_vary_coupl_strength_vibr_ZOFE}D, E (for $c=0.001$).
Figure~\ref{fig_vary_coupl_strength_vibr_ZOFE}B shows how for $c=0.1$ the excitation is transferred slightly faster from LHCII to CP43 and the RC than for the original coupling (shown as dashed lines for comparison), leading to the slightly larger amount of population reaching RP2 within 1~ns, observed in the bar diagram Figure~\ref{fig_vary_coupl_strength_vibr_ZOFE}A.
The population of pigment~7, which carries half of the initial population, shows oscillations now, indicating that part of the population is transported back and forth between pigments.
But this back-and-forth oscillation of a relatively small amount of population does not make the overall transport less efficient.
Also in the current from LHCII to CP43, shown in Figure~\ref{fig_vary_coupl_strength_vibr_ZOFE}C, there are now oscillations in the magnitude of the current. 
But the current still is always directed from the LHCII towards CP43, i.e., there is no backflow to LHCII, and the current is positive at all times.
On the other hand, the maxima of this current are more than twice as large as in the case of the original coupling to the vibrations.
(See Figure~\ref{fig_ZOFE_transfer}B for comparison, and note the different scales on the y axes).
The current from CP43 to the RC and from the RC to RP1 looks qualitatively similar to the original case.
There are no strong fluctuations in the magnitude of these currents.
But they are larger now compared to the original situation where the coupling to the vibrations is stronger.

The transport dynamics for the case of $c=0.001$ are shown in Figures~\ref{fig_vary_coupl_strength_vibr_ZOFE}D, E.
In panel~D, aside from the much smaller amount of population transported to RP2, we can also see that it takes much longer for the population to be transported away from the LHCII.
The reason for that becomes apparent when we look at the population of pigment~7 in LHCII:
the population oscillates back and forth between the LHCII pigments many times, and is not able to leave the LHCII.
It is trapped in the LHCII for a quite long time due to (unitary) localization inside the complex because of non-uniform pigment energies.
In panel~E, we see that the current between LHCII and CP43 oscillates fast between positive and negative values.
That means that excitation is flowing back and forth between these two proteins, quickly changing direction.
The other currents are almost zero, because only a small amount of population can pass through CP43 and reach the reaction center. 

Why the strong decrease of the vibrational coupling leads to such a trapping, can be understood with the notion of vibrationally enhanced transport:
the coupling to vibrations leads to transport through vibronic resonance and its absence leads to trapping.
This concept is depicted in Figure~\ref{vibronic_resonance}:
\begin{figure}
\centering
\includegraphics[width=0.8\mylenunit]{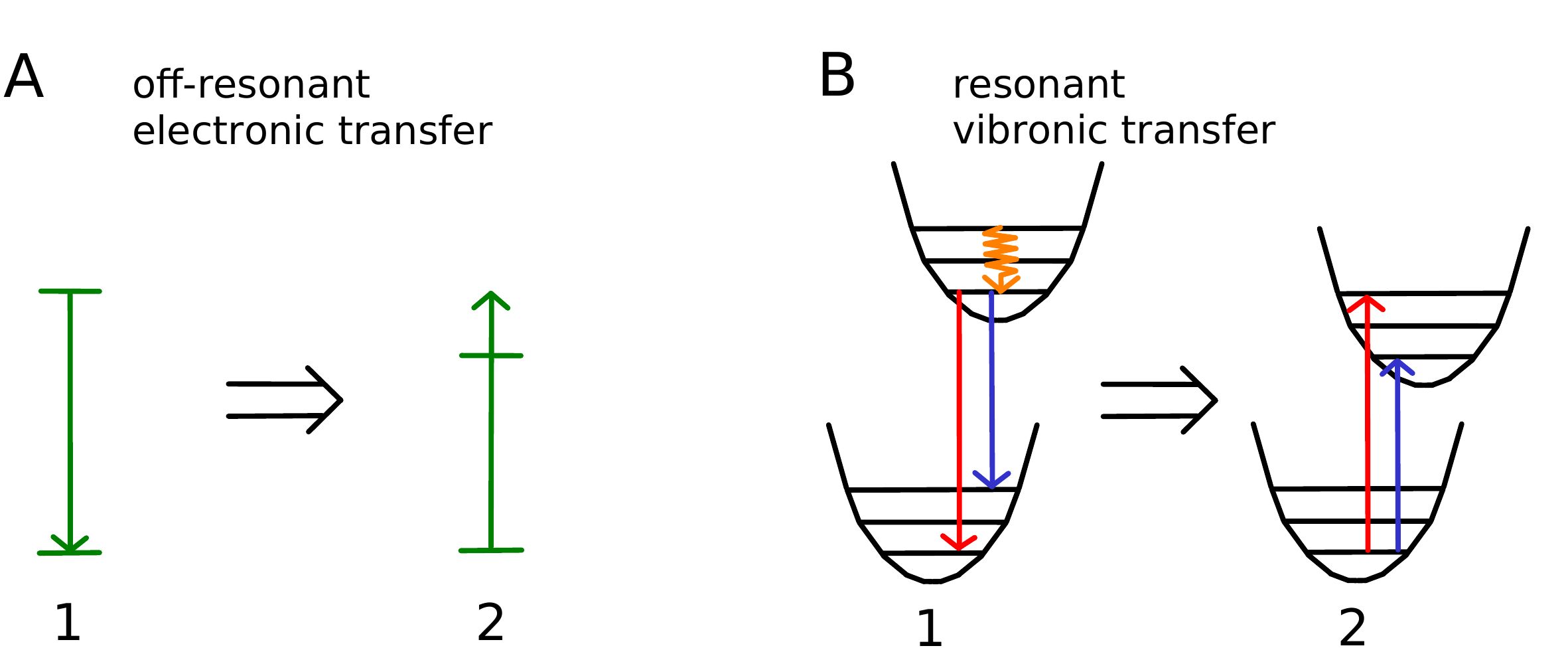}

\caption{Transport of electronic excitation from a pigment~1 to a pigment~2 via electronic inter-pigment interaction, when there is a gap between the pigment's electronic transition energies.
{\bf Figure~A:} without coupling to vibrations, the transport is off-resonant -- taking place between the off-resonant, purely electronic levels.
This off-resonance decreases the amount of population transferred to pigment~2.
{\bf Figure~B:} when there is coupling to vibrations -- with adequate strength and vibrational energies -- there can be resonance between the vibronic levels, leading to vibrationally enhanced transport.
(In this sketch the vibrational potential surfaces are only one-dimensional, i.e., they correspond to only one vibrational mode in each pigment, but the same argument applies for multiple modes, i.e., more-dimensional potential surfaces.
The environment spectral density can be decomposed into peaks of certain functional forms that represent damped molecular vibrational modes~\cite{roden2012accounting}.)
}
\label{vibronic_resonance}
\end{figure}
between the excited electronic states of the pigments there are energy gaps that, when they are large compared to the dipole-dipole interaction strength between the respective pigments, cause the excitation to be trapped on one or a few pigments, because of the off-resonance of the electronic transitions.
But if the electronic excitation of a pigment couples to vibrational modes with vibrational energies large enough and with a coupling strength large enough compared to the energy gaps, then the vibrational energy levels can fill the gaps and make the transport of excitation from one pigment to another resonant, thereby enhancing the transport.
Furthermore, the coupling to the continuum of vibrational modes leads to relaxation down the energy manifold that is needed for the energy funnel, directing the transfer to the RP states, to work.

For a delocalized initial state, in the case of very weak coupling to the vibrations, the trapping in the LHCII and CP43 will have smaller consequences for the tranfer efficiency, because there is also initial excitation in the reaction center that can be transferred to the RP states.
Further simulations have shown that for a delocalized initial condition the population that is transferred to RP2 -- in the case of the very weak coupling to the vibrations -- can be three times higher than for the localized initial state in Figure~\ref{fig_vary_coupl_strength_vibr_ZOFE}D.
However, the population transferred to RP2 is still by more than a factor of two lower than when the coupling to the vibrations has the original strength.
On the other hand, we have found in further simulations that for the original vibrational coupling $c = 1$, and when $c = 0.1$,~$0.01$, changing the initial state has little influence on the efficiency of the energy transfer. 
Only when the vibrational coupling is decreased beyond this, $c = 10^{-3}$--$10^{-4}$, does the initial state have a significant impact on the overall efficiency.
This shows again a high robustness of the energy transfer to variations of the initial condition.

\begin{figure}
\centering

\raisebox{0.05\mylenunit}{\includegraphics[width=0.5\mylenunit]{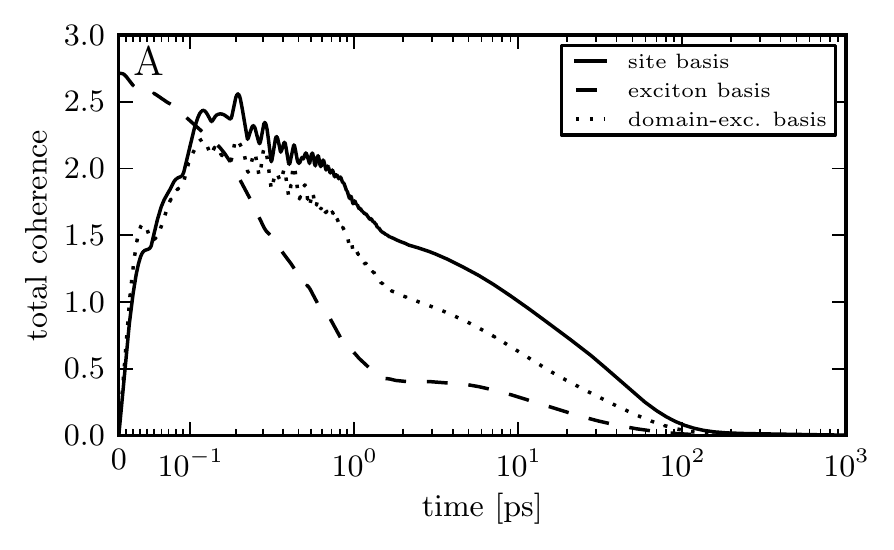}}
\includegraphics[width=0.45\mylenunit]{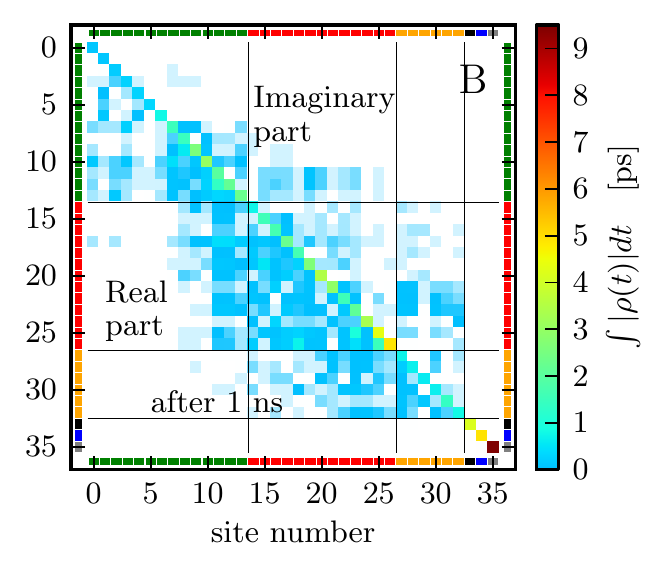}

\caption{Coherence from ZOFE simulation for weaker coupling to vibrations.
As Figures~\ref{fig_ZOFE_transfer}C, D, but the coupling to the vibrational environment is decreased {\bf by a factor of 10} compared to Figure~\ref{fig_ZOFE_transfer}.
}
\label{fig_ZOFE_transfer_coupl_times0.1_coher}

\end{figure}

Next, let us consider how decreasing the coupling to the vibrations affects the coherence involved in the transport dynamics. 
Figure~\ref{fig_ZOFE_transfer_coupl_times0.1_coher} shows the magnitude of coherence for the case of $c=0.1$ and should be compared to Figures~\ref{fig_ZOFE_transfer}C, D, for the original vibrational coupling.
For the weaker coupling to the vibrations (Figure~\ref{fig_ZOFE_transfer_coupl_times0.1_coher}), there is more coherence between the sites now, because the dephasing through the vibrations is weaker.
The excitonic coherence, by contrast, decreases (see panel~A compared to Figure~\ref{fig_ZOFE_transfer}C) which is expected because part of it is created by the coupling to the vibrations.   
Remarkably, these changes in the coherence do not affect the overall transport efficiency much, as we have seen in Figure~\ref{fig_vary_coupl_strength_vibr_ZOFE}.
%
%

\subsection{Full Markovian Lindblad simulation: larger range of coupling strength}
\label{sec_full_markov_lindblad_simul}

So far, we have only {\it decreased} the coupling to the vibrations in relation to our original model.
It would also be instructive to investigate the energy tranfer at stronger coupling as well.
Stronger coupling, however, leads to problems in the ZOFE calculation, because of the limited range of non-Markovianity for which the ZOFE approximation is valid~\cite{strunz2004convolutionless, yu1999_91, roden2011zofe_test_spectra}.
To still be able to gain an insight into the behavior at stronger coupling and -- at least roughly -- assess the energy transfer efficiency, we therefore use a full Markovian Lindblad calculation, where the Markov approximation is applied to the electron-vibration coupling.
Based on the reasoning of Ref.~\cite{jesenko2013_174103}, an assessment of the transport efficiency can be reasonable with an appropriate Markovian description. 
As a check, we also calculate the cases of weaker vibrational coupling again with the Markovian Lindblad simulation, so that we can compare with the previous ZOFE simulation results.
A description of the Lindblad calculation is given in Appendix~\ref{app_pure_lindblad_simul}.

Figure~\ref{fig_Lindblad_vary_coupl_strength_vibr} summarizes the effect of stronger couplings with such a Lindblad calculation, where the strength of the coupling to the vibrational environment is varied by a factor between 0.0001 and 50.
Here, factor 1 corresponds to the coupling strength in the original model.
\begin{figure}
\centering

\includegraphics[width=0.54\mylenunit]{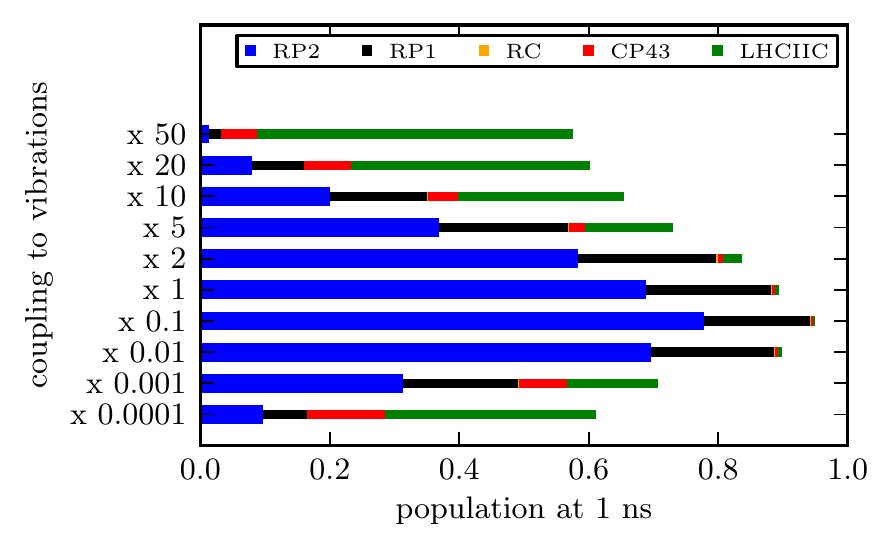}

\caption{Varying the strength of the coupling to vibrational environment; fully Markovian Lindblad simulation.
The strength of the coupling to the vibrations is varied by a factor between 0.0001 and 50. 
(Factor 1 corresponds to the original model.)
The populations resulting from the Lindblad simulation are shown (in the same way as in Figure~\ref{fig_vary_coupl_strength_vibr_ZOFE}A).
The initial state is the same localized initial state as for Figure~\ref{fig_vary_coupl_strength_vibr_ZOFE}, where all population is on the two pigments~7 and~10 in the LHCII, without coherence in the site basis.}
\label{fig_Lindblad_vary_coupl_strength_vibr}
\end{figure}
We see that as before for the ZOFE calculation, the Lindblad results also show the highest transport efficiency in the case where the vibrational coupling is decreased by a factor of ten with respect to the original model.
In this case, also the amount of integrated population of the RP2 state is very similar to the ZOFE result.
However, the RP2 populations for the original coupling and for the cases of weakest coupling, calculated with Lindblad, are in less good agreement with the ZOFE results.

Let us now look at the cases where the vibrational coupling is increased compared to the original model. 
In Figure~\ref{fig_Lindblad_vary_coupl_strength_vibr}, we see that as the coupling is increased, the transport efficiency becomes lower and lower, until for a fifty-fold increase, it is almost zero.
At the same time the amount of population that stays in the LHCII becomes larger and larger.
This suppression of the transport when the coupling to the vibrations is stronger may be understood in terms of the quantum Zeno effect~\cite{rebentrost2009_033003}, where the electronic excitation interacts with the vibrations -- i.e., it is ``measured'' by the vibrations in the site basis -- so strongly, that is, at such a high rate, that the excitation is forced to stay in the original state and cannot move anymore. 
In terms of the expression for the population currents, Equation~(\ref{total_current}), this effect can be understood from the fact that the stronger coupling to the vibrations causes stronger dephasing, destroying the coherence between the sites that drives the energy transfer, and thereby diminishes the energy transfer.
Because of this effect at stronger coupling and the trapping of the excitation at weak coupling, the highest transport efficiency is obtained in the intermediate regime.
There, the magnitude of the coupling to the vibrations is such that it gives the optimal balance between vibrationally enhanced transport (as described in Figure~\ref{vibronic_resonance}) and the quantum Zeno effect.
These results of the Lindblad calculation therefore provide more evidence that the strength of the coupling to the vibrations that we used in our original model lie roughly in the region of optimal coupling strength.
However, remarkably, these results again show that the energy transfer efficiency could be even (slightly) higher at a factor $\sim 10$ weaker coupling to the vibrations.

\section{Summary and conclusions}
\label{sec_conclusion}

In this work, we simulated the long-range inter-complex energy transfer in the PSII supercomplex -- from LHCII via the core complex CP43 to the reaction center -- by means of a full quantum calculation, using a non-Markovian ZOFE quantum master equation description.
We found that nearly all energy from the initial electronic excitation is transported to the reaction center within about 200~ps, and that after 1~ns, about 80\% of the energy is transformed into charge separation in the reaction center.
Less than 5\% is lost through radiative and non-radiative decay within this time. 

These findings for the overall energy transfer dynamics are in good agreement with the results of Ref.~\cite{bennett2013_9164}, where a modified-Redfield-generalized-F\"{o}rster (MRGF) rate-equation method was applied to calculate the long-range energy transfer in the entire PSII supercomplex. 
In the MRGF method, strongly coupled pigments are assigned to domains treated with a modified-Redfield description, whereas excitation transfer between the domains, where the couplings are weaker, is treated with a generalized-F\"{o}rster description.
In the present work, to compare our ZOFE quantum-master-equation results to the MRGF rate-equation description of Ref.~\cite{bennett2013_9164}, we used the MRGF method to calculate the energy transfer in the truncated supercomplex LHCII-CP43-RC of PSII that we considered. 
The resulting energy transfer dynamics are in very good quantitative agreement with that calculated with the ZOFE master equation, showing that the simpler MRGF rate equation provides an adequate description of the quantum energy transfer dynamics and energy transfer efficiency of PSII. 
The observation that the MRGF results agree well with the ZOFE results shows that the subdivision into domains, which is related to the notion of supertransfer, i.e., coherent delocalization inside domains leads to enhanced transfer rates between the domains~\cite{kassal2013_362}, is a reasonable approach to describe energy transfer in large light-harvesting complexes, where the coupling between the pigments can be very inhomogeneous. 
We have further shown that excitation energy transfer is significantly slower within a pure F\"{o}rster calculation, which further supports the importance of supertransfer for rapid excitation transport within PSII. 
The close agreement between ZOFE and MRGF is also consistent with the findings of Ref.~\cite{jesenko2013_174103}, where a full quantum description of electronic excitation transfer is compared to that of a classical rate equation, and it is shown that the rate equation description is capable of capturing the overall features of the energy transfer when properly derived from the full quantum description.

In addition to the commonly used description in terms of time-dependent populations of the excited electronic states of the pigments, we also analyzed the energy transfer dynamics in terms of excitation population {\it currents} between the pigments. 
This analysis revealed the pathways and directions of the energy transfer between the individual pigments and the net fow of energy between the different complexes.
We found that when initially all excitation is in the LHCII antenna complex, the net flow of energy is directed towards the reaction center at all times and no temporary intercomplex backflow occurs.
Integrating the population currents over time showed that there is a complex network of pathways rather than just one or two dominant pathways, in accordance with the findings of Ref.~\cite{bennett2013_9164}.
The energy transfer between different proteins was found to rely on several different pathways.
Interestingly, instead of being uniformly directed towards the reaction center, often excitation energy is transported in loops between pigments, i.e., excitation is transferred from one pigment to a neighboring pigment and from there to the next, etc., in such a way that it goes around in a closed circuit of pigments and in the end comes back to the initial pigment.
Such loops were found in all three subcomplexes that we considered, LHCII, CP43, and the reaction center.
In the reaction center, the efficient excitation energy transport to the charge separated states appeared to rely mainly on only two pigments (30 and 32), providing evidence that the rest of the reaction center pigments are needed primarily for their role in the electron transport rather than for the excitation transfer; 
some of these reaction center pigments appear to constitute a strong loop excitation current, but do not contribute to the net transport of excitation energy to the charge separated states. 
To test this conclusion, which was drawn from the analysis of the pathways of integrated population currents, we ran additional simulations with all reaction center pigments -- except for 30 and 32 -- decoupled from the rest of the pigments so that transfer of excitation to and from these decoupled pigments was inhibited, but leaving the coupling to the charge separated states unmodified. 
The resulting efficiency of the energy transport to the charge separated states in this restricted model was found to be the same as in the original model where excitation transport to the other reaction center pigments was allowed, showing that the excitation energy transport to and inside the reaction center appears to rely indeed only on the two pigments 30 and 32. 
This confirmed the conclusions drawn from the analysis of the integrated population currents and showed that this analysis is suited to identify transport pathways and assess their importance, providing a useful tool for the investigation of design-function relationships in light-harvesting systems.

By simulating the energy transfer with the ZOFE master equation, we were able to quantify the amount of electronic coherence that is involved in the energy transfer dynamics. 
Motivated by the description of the energy transfer in terms of population currents and the different roles of imaginary and real part of the coherence presented in a companion paper~\cite{roden2015prob_current}, we quantified the real and imaginary components of the coherence between the pigments separately.
We found that the real components, which are related to localization of excitation due to energy gaps between the pigments~\cite{roden2015prob_current}, are overall larger than the imaginary components, which are proportional to the population currents between the pigments and drive the excitation energy transfer in the PSII supercomplex.

To find out how the energy transfer in PSII depends on the initial condition -- which is determined by the specific features of the light used for the excitation -- we ran simulations of the energy transfer for different initial states;
first we considered initial excitation that is localized on only two LHCII pigments;
then we started from an initial state in which the excitation is completely delocalized over the LHCII, CP43, and reaction center pigments.
We found that even though for the delocalized initial state the energy transfer dynamics of course look different from that for the localized initial state, the overall efficiency of the energy transfer to the charge separation in the reaction center is nevertheless the same in both cases.
Also the overall pattern of the energy transfer pathways between the individual pigments is very similar for the different initial conditions -- for both short-range (intra-complex) and long-range (inter-complex) energy transfer.
This shows that the energy transfer in PSII is remarkably robust with respect to the initial conditions.

To learn how sensitive the energy transfer dynamics are to variations in the coupling between the electronic and vibrational degrees of freedom, we varied the strength of this coupling to the vibrations in our simulations:
decreasing the coupling to the vibrations for all pigments by a factor of ten compared to the original situation did not change the energy transfer efficiency much -- it even slightly increased the efficiency, showing that the energy transfer is very robust to variations of the coupling to the vibrations.
Even when the coupling to the vibrations was decreased by a factor of 100, the amount of energy transferred to the charge separation within 1~ns decreased only by about 10\%.
We assign this to a decrease of the well known effect of vibrationally enhanced transport, which helps the transfer of energy through more resonant energy levels when coupling to vibrations (vibrational energy levels) is present.  
Only when the electron-vibration coupling was decreased by a factor of 1,000, does the amount of energy transferred decrease significantly (by up to a factor 4 compared to the original situation), due to trapping of the excitation in the LHCII complex.
This trapping is characteristic of the predominantly unitary dynamics occuring in the absence of vibrationally enhanced transport.
These observations show that for PSII the strength of the coupling to the vibrations that is needed for the effect of vibrational enhanced transport to direct the energy transfer efficiently to the reaction center is rather small.

However, in contrast to the Fenna-Matthews-Olson (FMO) complex, where calculations have revealed that the transport efficiency does not fall under 70\% when the coupling to the vibrational environment is decreased, even to extremely small coupling strength~\cite{rebentrost2009_9942}, for PSII we found that the efficiency can become very low -- almost zero -- for extremely small coupling to the vibrations.  
This shows that contrary to FMO, for PSII the coupling to the vibrations is crucial for achieving efficient transport.

To assess how the energy transfer changes when the coupling to the vibrations is increased compared to the original situation, we performed Markovian Lindblad simulations, instead of the ZOFE calculations because the latter were not applicable in this parameter range.
Here, we found that if the coupling to the vibrations is stronger than in the original situation and is further increased step by step, the efficiency of the energy transfer decreases monotonically, until for a fifty-fold increase of the coupling, almost no energy is transferred to the reaction center anymore (within the reference time of 1~ns) due to the strong dephasing that destroys the coherence needed for the energy transfer~\cite{roden2015prob_current}.  
These approximate Lindblad simulations, together with the ZOFE simulations, thus provide evidence that the original strength of the coupling to the vibrations lies in the parameter region for optimal energy transfer efficiency. 
We showed, however, that for a coupling that is about ten times weaker than in the original situation, the efficiency could be even slightly higher.
This could be an important feature to reflect upon, in the light of the discussion about the possible (evolutionary) optimization of natural light-harvesting systems like PSII.

\appendix

\section{Pure Lindblad simulation of excitation energy transfer}
\label{app_pure_lindblad_simul}

To be able to roughly assess the efficiency of excitation energy transfer in our truncated PSII model for an extended range of the electron-vibration coupling, in Section~\ref{sec_full_markov_lindblad_simul} we used a pure Lindblad simulation for the excitation energy transfer, applying the Markov approximation to the electron-vibration coupling.
Here, we briefly describe the calculation and how the corresponding parameters are obtained.

For the simulation, we employ a Lindblad master equation~(\ref{lindblad_master}), where we now include Lindblad terms describing the coupling of the electronic excitation to the vibrations of the pigments and protein environment -- in addition to the Lindblad terms for transfer to the RP1 and RP2 states and radiative and non-radiative decay to the ground state (relaxation terms) that were also included in the ZOFE simulations.
The new Lindblad terms describing the electron-vibrational coupling contain Lindblad operators $L_n = \ket{n}\bra{n}$ and coupling parameters $\gamma^{\rm Lind}_n$ for each pigment $n$, so that -- instead of Eq.~(\ref{zofe_lindblad_master}) for the ZOFE simulation -- we now have the pure Lindblad master equation
\begin{equation}
\label{full_lindblad_master}
\begin{split}
  \partial_t \rho(t) = -i\left[H_{\rm sys}, \rho\right] & + \sum_{n=0}^{32}\gamma^{\rm Lind}_n\left(L_n\rho L_n^{\dagger} - \frac{1}{2} L_n^{\dagger}L_n\rho - \frac{1}{2}\rho L_n^{\dagger}L_n \right)\\
&  + \sum_{i=0}^{39}\widetilde{\gamma}_i\left(\widetilde{L}_i\rho \widetilde{L}_i^{\dagger} - \frac{1}{2}\widetilde{L}_i^{\dagger}\widetilde{L}_i\rho - \frac{1}{2}\rho \widetilde{L}_i^{\dagger}\widetilde{L}_i \right),
\end{split}
\end{equation}
where the terms in the second line again describe transfer to the RP1 and RP2 states and radiative and non-radiative decay.
In the Lindblad term in the first line, the electronic excitation of each pigment $n$ is coupled to vibrations with respective coupling strength $\gamma^{\rm Lind}_n$. 
To calculate the excitation energy transfer from the full Lindblad description~(\ref{full_lindblad_master}), we need to assume values for the coupling parameters $\gamma^{\rm Lind}_n$.
In the following, we show how we obtain these parameters through a simple approximation from the parameters that we used to describe the non-Markovian electron-vibration coupling within the ZOFE description.  
To this end, let us again consider the special form of the environment correlation function~(\ref{alpha_sum_exps}) used in the ZOFE description
\begin{equation}
\label{APP_alpha_sum_exps}
  \alpha_n(t-s) = \sum_j\Gamma_{nj}e^{-i\Omega_{nj}(t-s)}e^{-\gamma_{nj}|t-s|}.
\end{equation} 
The pure Lindblad description~(\ref{full_lindblad_master}) is obtained in the Markov limit, where 
\begin{equation}
\label{markov_alpha_is_delta}
  \alpha_n(t-s) = \gamma^{\rm Lind}_n \delta(t-s),
\end{equation} 
i.e.\ $\alpha_n(t-s)$ decays fast compared to the timescales of the system dynamics.
To express the parameters $\gamma^{\rm Lind}_n$ through the parameters of Eq.~(\ref{APP_alpha_sum_exps}) in this limit, we consider the Dirac series $\delta_{\epsilon}(x) = \frac{1}{2\epsilon}\exp(-|x|/\epsilon)$ for $\epsilon\rightarrow 0$, which lets us write
\begin{equation}
  \label{limit_dirac_series}
  e^{-\gamma_{nj}|t-s|} = \frac{2}{\gamma_{nj}}\,\delta(t-s)\ \ \ \mbox{for} \ \ \ \gamma_{nj}\rightarrow\infty.
\end{equation}
Inserting Eq.~(\ref{limit_dirac_series}) in Eq.~(\ref{APP_alpha_sum_exps}) we have
\begin{equation}
\label{alpha_dirac_series}
  \alpha_n(t-s) = \left(\sum_j\Gamma_{nj}e^{-i\Omega_{nj}(t-s)}\frac{2}{\gamma_{nj}}\right)\,\delta(t-s)\ \ \ \mbox{for} \ \ \ \gamma_{nj}\rightarrow\infty.
\end{equation} 
Inserting~(\ref{alpha_dirac_series}) in the Eq.~(\ref{o_bar}) for the ZOFE auxiliary operator, we get
\begin{equation}
  \label{o_bar_markov_limit}
  \oop{(n)}(t) = \frac{1}{2}\sum_j\Gamma_{nj}e^{-i\Omega_{nj}(t-t)}\frac{2}{\gamma_{nj}} O^{(n)}_0(t,t) = \sum_j\frac{\Gamma_{nj}}{\gamma_{nj}}L_n\ \ \ \mbox{for} \ \ \ \gamma_{nj}\rightarrow\infty,
\end{equation}
since $O^{(n)}_0(t,t)=L_n$ (see Eq.~(\ref{o_init})).
Since we know that in the Markov limit $\alpha_n(t-s) = \gamma^{\rm Lind}_n \delta(t-s)$, the auxiliary operator of the ZOFE master equation becomes the constant operator $\oop{(n)}(t)=\frac{1}{2}\gamma^{\rm Lind}_n L_n$ (see Eqs.~(\ref{o_bar}) and (\ref{o_init}))~\cite{strunz2004convolutionless}, from comparison with Eq.~(\ref{o_bar_markov_limit}) it follows that we can approximate the needed coupling parameters for the full Lindblad descriptions as
\begin{equation}
  \label{gamma_lindblad}
  \gamma^{\rm Lind}_n = 2\sum_j\frac{\Gamma_{nj}}{\gamma_{nj}}\ \ \ \mbox{for} \ \ \ \gamma_{nj}\rightarrow\infty,
\end{equation}
i.e.\ in the case where the $\gamma_{nj}$ are large compared to the system energies. 
We note that in order for the $\gamma^{\rm Lind}_n$ to stay finite in the $\gamma_{nj}\rightarrow\infty$ limit, not only the $\gamma_{nj}$, but also at least one of the parameters $\Gamma_{nj}$ has to go to infinity.
Inserting the parameters for the electron-vibration coupling of the ZOFE simulation (according to Section~\ref{sec_approx_spec_dens_for_zofe}) into Eq.~(\ref{gamma_lindblad}), we find the following values for the parameters $\gamma^{\rm Lind}_n$ that we used for our full Lindblad simulation in Section~\ref{sec_full_markov_lindblad_simul}:\\ 
$\gamma^{\rm Lind} \approx 4,200\ {\rm cm}^{-1}$ for the Chl~A pigments in LHCII, \\
$\gamma^{\rm Lind} \approx 5,280\ {\rm cm}^{-1}$ for the Chl~B pigments in LHCII, \\
$\gamma^{\rm Lind} \approx 140\ {\rm cm}^{-1}$ for the pigments in CP43, \\
and $\gamma^{\rm Lind} \approx 190\ {\rm cm}^{-1}$ for the pigments in the reaction center.\\
We can see from the fact that these values are not at all large compared to the system energies (the differences of the eigenenergies of the system Hamiltonian are in the range of up to 1,300~cm$^{-1}$) that the Markov approximation is not well justified here and that therefore our full Lindblad simulations should be considered to be a rather crude approximation of the non-Markovian dynamics.

\section{Effect of coupling to the radical pair states}

To see the influence of the radical pair states on the transport dynamics, we now look at the dynamics with the coupling to the RP states switched off.
That is, there is no transfer of population to the radical pair states.

\begin{figure}
\centering
\hspace{-0.02\mylenunit}\includegraphics[width=0.5\mylenunit]{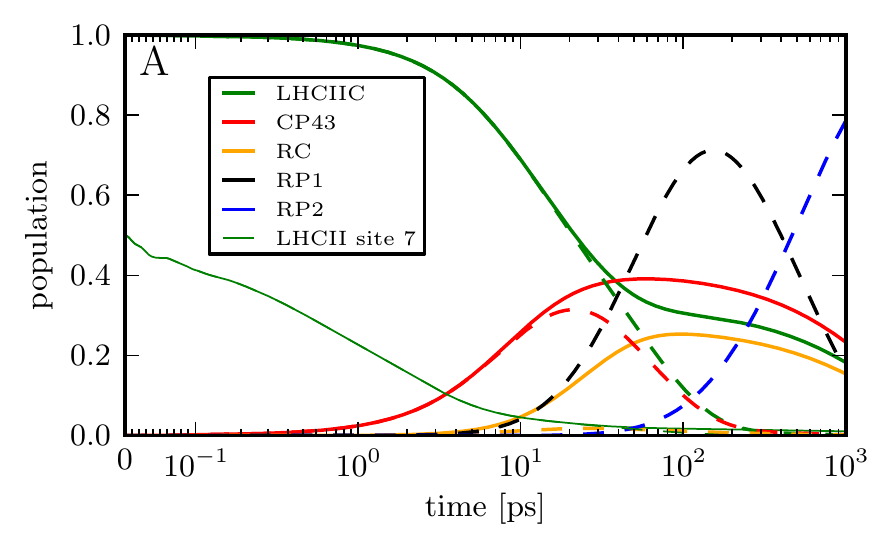}\hspace{-0.02\mylenunit}
\includegraphics[width=0.5\mylenunit]{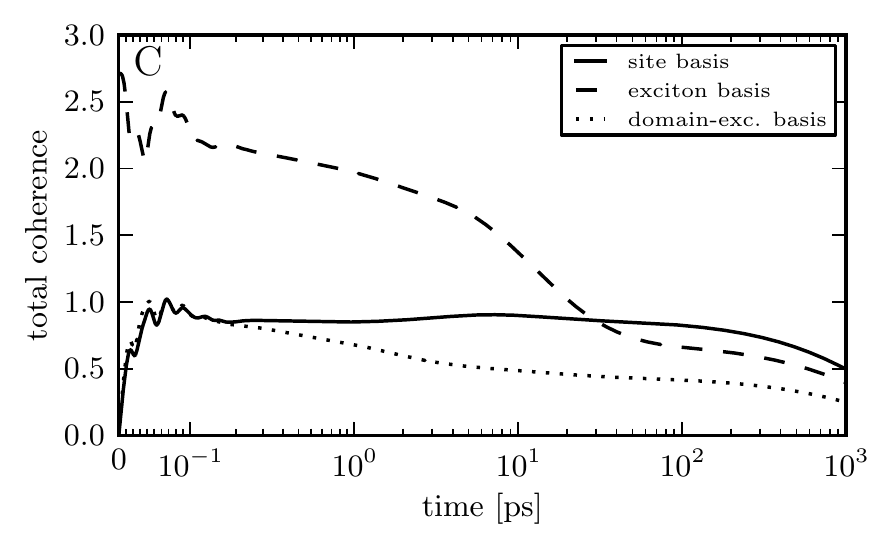}

\caption{As Figure~\ref{fig_ZOFE_transfer}A and~C, but no coupling to the RP1 and RP2 states.
For comparison, the dashed lines in Figure~A show the ZOFE dynamics from Figure~\ref{fig_ZOFE_transfer}A (solid lines in Figure~\ref{fig_ZOFE_transfer}A) where the coupling to the RP states is present.
}
\label{fig_ZOFE_transfer_noCoupltoRP}
\end{figure}

Figure~\ref{fig_ZOFE_transfer_noCoupltoRP} shows the resulting dynamics for the same quantities as in Figure~\ref{fig_ZOFE_transfer}, but with the coupling to the RP states switched off.
We see in Figure~\ref{fig_ZOFE_transfer_noCoupltoRP}A that up to a certain time, the populations in the proteins with and without coupling are equal.
The further a protein is away from the RP states, the later the population without the coupling deviates from the population where the coupling is present.
Instead of falling off fast, now -- without the coupling to the RP states -- after around 200~ps the populations decline much slower;
this residual, slower decline is due to the radiative and non-radiative decay of the pigment excited states. 
After 200~ps, the residual, average population per pigment is highest in the RC, and it is lowest in LHCII.
This is because the average energy in the RC is lowest and that in LHCII is highest, providing the ``energy funnel'' that directs the transport to the RC.

Let us next look at the coherence.
When we compare the total coherence in Figure~\ref{fig_ZOFE_transfer_noCoupltoRP}C, where the coupling to the RP states is switched off, with that in Figure~\ref{fig_ZOFE_transfer}C, where the coupling to the RP states is present, we see that up to about 10~ps the coherence with and without coupling is equal.
After that, however, the coherence without coupling to the RP states goes to finite values that decrease only slowly, whereas in the case when the coupling is present, the coherence falls off to zero. 
This behavior can be explained with the Cauchy-Schwarz inequality 
\begin{equation}
  \label{cauchy_schwarz}
  |\rho_{ij}(t)|^2 \leq \rho_{ii}(t)\,\rho_{jj}(t) 
\end{equation}
that holds for the coherences $\rho_{ij}(t)$ and the populations $\rho_{ii}(t)$, $\rho_{jj}(t)$ of any density matrix $\rho(t)$.
It says that coherences can only ever arise between states that are populated. 
The larger the populations of two given states are, the larger can the coherence between these states potentially be.
In the case where there is transfer to the RP states, population is taken away from the excited states of the pigments by the RP states, and consequently, the coherence, being limited by the population through the Cauchy-Schwarz inequality, has to fall off to zero.
On the other hand, when the transfer to the RP states is switched off, the population remains in the excited states of the pigments and the coherence stays finite.

\begin{acknowledgments}

This material was supported by DARPA under Award No.\ N66001-09-1-2026.
J.~R.\ thanks Brendan Abolins, Daniel Freeman, Siva Darbha, Jon Aytac, Ty Volkoff, Felix Motzoi, Loren Greenman, Donghyun Lee, and Aleksey Kocherzhenko for helpful discussions.

\end{acknowledgments}




\end{document}